\documentclass[epsfig,12pt]{article}
\usepackage{makeidx}
\usepackage{amsmath}
\usepackage{amsfonts}
\usepackage{amssymb}
\usepackage{graphicx}

\input epsf.sty
\makeatletter \@addtoreset{equation}{section}
\renewcommand{\theequation}{\thesection.\arabic{equation}}
\input{tcilatex}
\begin{document}

\title{\vspace{-2cm}%
\rightline{\mbox{\small
{\bf LPHE-MS-10-03  /  CPM-1002}}} \bigskip \bigskip \textbf{Intersecting
Black Attractors in }\\
\textbf{\emph{8D } }$\mathcal{N}\mathbf{=1}$\textbf{\ Supergravity}\bigskip }
\author{R. Ahl Laamara$^{1,2}$, L.B Drissi$^{3}$, F.Z Hassani$^{1}$, E.H
Saidi$^{2}$, A.A Soumail$^{1,2}$\bigskip \\
%EndAName
{\small 1. LPHE-Modeling and Simulation, Facult\'{e} des Sciences, Rabat,
Morocco}\\
{\small 2. Centre of Physics and Mathematics, CPM, CNESTEN, Morocco}\\
{\small 3. Inanotech-MAScIR, Institute of Nanomaterials and Nanotechnology,
Rabat, Morocco}}
\maketitle

\begin{abstract}
We study intersecting extremal black attractors in non chiral \emph{8D} $%
\mathcal{N}=1$ supergravity with moduli space $\frac{SO\left( 2,N\right) }{%
SO\left( 2\right) \times SO\left( N\right) }\times SO\left( 1,1\right) $ and
work out explicitly the attractor mechanism for various black p-brane
configurations with the typical near horizon geometries $AdS_{p+2}\times
S^{m}\times T^{6-p-m}$. We also give the classification of the solutions of
the attractor equations in terms of the $SO\left( N-k\right) $ subgroups of $%
SO\left( 2\right) \times SO\left( N\right) $ symmetry of the moduli space as
well as their interpretations in terms of both heterotic string on 2-torus
and its type IIA dual. Other features such as non trivial SO$\left(
1,7\right) $ central charges $Z_{\mu _{1}...\mu _{p}}$ in \emph{8D} $%
\mathcal{N}=1$ supergravity and their connections to p-form gauge fields are
also given.\newline
\end{abstract}

\section{Introduction}

During the last decade, black attractor solutions in supergravity
theories have been a subject of big interest; especially in
connection with low energy \emph{10D} superstring and \emph{11D}
M-theory compactifications
\textrm{\cite{1A}-\cite{2F}}. Because of their specific properties \textrm{%
\cite{3A}-\cite{4D}}, static, asymptotically flat and spherically symmetric
extremal (vanishing temperature for non-zero entropy) black attractors have
been investigated for supergravities in diverse space time dimensions; with
various numbers of conserved supersymmetries \textrm{\cite{5A}-\cite{ss}}.%
\newline
Guided by the new solutions on extremal BPS and non BPS black attractors in
higher dimensional supergravity; in particular those on intersecting
attractors obtained first by Ferrara et al. in \textrm{\cite{70}}, see also
\textrm{\cite{7A}}; we focus in this paper on non chiral \emph{8D} $\mathcal{%
N}=1$ supergravity with moduli space $\frac{SO\left( 2,N\right) }{SO\left(
2\right) \times SO\left( N\right) }\times SO\left( 1,1\right) $ and study
explicitly the attractor mechanism for various configurations of extremal
black p-branes with the typical near horizon geometries $AdS_{p+2}\times
S^{m}\times T^{6-p-m}$ where $p=0,1,2,3,4$ and $m=2,3,4,5,6$. Actually this
analysis completes the results obtained in \textrm{\cite{7A}} for the case
of \emph{maximal} $\mathcal{N}=2$ supergravity in \emph{8D}; it also gives
new solutions, along the line of \textrm{\cite{70}}, classified by $SO\left(
N-k\right) $ subgroups of the $SO\left( 2\right) \times SO\left( N\right) $
symmetry of the moduli space of the non chiral \emph{8D} $\mathcal{N}=1$
supergravity. \newline
The interest into this study is also motivated from the two following
features: first because of its \emph{16} conserved supersymmetries, extremal
black attractors in this \emph{8D} supergravity may be viewed as the
ancestor of an interesting class of black holes in \emph{7D,} \emph{6D,}
\emph{5D }and \emph{4D} supergravities \textrm{\cite{1A,9A}}; in particular
in \emph{4D} $\mathcal{N}=2$ and \emph{4D} $\mathcal{N}=4$ resulting from
adequate compactifications of the \emph{8D} space time down to \emph{4D}. It
is also interesting from the view of higher dimensions since non chiral
\emph{8D} $\mathcal{N}=1$ supergravity may arise as low energy of heterotic
string on $T^{2}$ and type IIA string on a real compact surface $\Sigma $
that preserves half of the \emph{32} conserved supercharges of the \emph{10D}
type II superstrings. Black attractor solutions in \emph{8D} offers
therefore a framework to explicitly check specific features of the
heterotic/type IIA duality in \emph{8D }\textrm{\cite{100}}. \newline
The paper is organized as follows.\textit{\ }\textrm{In section 2}, we
review the \emph{8D} $\mathcal{N}=1$ supersymmetry algebra with central
charges $Z_{\mu _{1}...\mu _{p}}$. We derive the various $SO\left(
1,7\right) $ charges of these $Z_{\mu _{1}...\mu _{p}}$'s and give their
connection with p-branes. It is also shown why the 3-form gauge field in $%
\mathcal{N}=1$ theory should vanish. \textrm{In section 3}, we develop the
study of the non chiral $\mathcal{N}=1$ supergravity in \emph{8D} and its
links with the low energy limit of the heterotic superstring on T$^{2}$ and
its type $IIA$ superstring dual. The various charges of the black attractors
are also given\textbf{.} \textrm{In section 4}, we first give the effective
potential and the attractor eqs; then we derive their solutions together
with their classification in terms of $SO\left( N-k\right) $ subgroups of
the $SO\left( 2\right) \times SO\left( N\right) $ symmetry of the moduli
space.\textit{\ }\textrm{In section 5, }we study the intersecting attractors
along the line of the approach of \textrm{\cite{70,7A}} and \textrm{in
section 6}, we give a conclusion.

\section{Central charges in \emph{8D} $\mathcal{N}=1$ supersymmetry}

In this section, we identify the full set of the bosonic "central charges" $%
Z_{\mu _{1}...\mu _{p}}$ involved in the generalized non chiral \emph{8D} $%
\mathcal{N}=1$ superalgebra and give their links to black p-brane attractors
by using group theoretical methods.\newline
To start it is interesting to recall that like in \emph{4D} space time,
supersymmetry with sixteen supercharges may also live in other space times.\
In eight dimensions; this is precisely $\mathcal{N}=1$ supersymmetry given
by a graded superalgebra with both commutators and anticommutators; it
exchanges \emph{8D} space time bosons into \emph{8D} space time fermions. In
addition to the twenty eight $M_{\left[ \mu \nu \right] }$ symmetry
generators of $SO\left( 1,7\right) $, the standard (non extended) $\mathcal{N%
}=1$ non chiral supersymmetry is moreover generated by the energy momentum
vector operator $P_{\mu }$ and the fermionic generators $Q_{\alpha }$ and $%
\bar{Q}_{\dot{\alpha}}$ transforming as Weyl spinors under $SO\left(
1,7\right) $. To have more insight on the structure of this superalgebra and
its connection with black branes in \emph{8D}, we give below some useful
details.

\subsection{$\mathcal{N}=1$ superalgebras in \emph{8D}}

We begin by recalling the group theoretical nature of the fermionic
generators in \emph{8D} $\mathcal{N}=1$ non chiral supersymmetry; these are $%
SO\left( 1,7\right) $ spinors with $2^{4}=16$ complex components that
transform in the reducible $8_{s}\oplus 8_{c}$ representation of the \emph{8D%
} Lorentz group respectively given by the eight component Weyl spinors $%
Q_{\alpha }^{+}$ and $\bar{Q}_{\dot{\alpha}}^{-}$. These fermionic
generators carry also charges under the $SO\left( 2\right) \sim U_{R}\left(
1\right) $ R-symmetry of the supersymmetric algebra.\newline
Using general properties of tensor products of $SO\left( 1,7\right) \times
U_{R}\left( 1\right) $ representations, we learn that one may distinguish
three kinds of $\mathcal{N}=1$ anticommutation relations in \emph{8D}; two
chiral (complex) relations and a vector like (real) one:

$\left( 1,0\right) $\emph{\ chiral relations in 8D}\newline
These are complex relations involving only the fermionic generator $%
Q_{\alpha }^{+}$,%
\begin{equation}
\begin{tabular}{llll}
$\left\{ Q_{\alpha }^{+},Q_{\beta }^{+}\right\} =Z_{\left( \alpha \beta
\right) }^{++}$ & , & $\left[ Q_{\alpha }^{+},Q_{\beta }^{+}\right] =Z_{%
\left[ \alpha \beta \right] }^{++}$ & ,%
\end{tabular}
\label{q}
\end{equation}%
where the symmetric $Z_{\left( \alpha \beta \right) }^{++}$'s should be
thought of as operators carrying charges of black branes in \emph{8D}. The
antisymmetric term $Z_{\left[ \alpha \beta \right] }^{++}$'s, which may be
expanded as $\sigma _{\left[ \alpha \beta \right] }^{\mu \nu }M_{\mu \nu }$
may be interpreted in terms of $SO\left( 1,7\right) $ rotations in the Weyl
representation.

$\left( 0,1\right) $\emph{\ antichiral relations in 8D}\newline
These are the complex conjugate of (\ref{q});\emph{\ }they involve the $\bar{%
Q}_{\dot{\alpha}}$ spinor%
\begin{equation}
\begin{tabular}{llll}
$\left\{ \bar{Q}_{\dot{\alpha}}^{-},\bar{Q}_{\dot{\beta}}^{-}\right\} =\bar{Z%
}_{\left( \dot{\alpha}\dot{\beta}\right) }^{--}$ & , & $\left[ \bar{Q}_{\dot{%
\alpha}}^{-},\bar{Q}_{\dot{\beta}}^{-}\right] =\bar{Z}_{\left[ \dot{\alpha}%
\dot{\beta}\right] }^{--}$ & ,%
\end{tabular}%
\end{equation}%
where $\bar{Z}_{\left( \dot{\alpha}\dot{\beta}\right) }^{--}$ and $\bar{Z}_{%
\left[ \dot{\alpha}\dot{\beta}\right] }^{--}$ are the complex conjugate of $%
Z_{\left( \alpha \beta \right) }^{++}$ and $Z_{\left[ \alpha \beta \right]
}^{++}$.

$\mathcal{N}=\left( 1,1\right) $\emph{\ superalgebra in 8D}\newline
This is a \emph{vector like }superalgebra with fermionic generators $%
Q_{\alpha }^{+}$ and $\bar{Q}_{\dot{\alpha}}^{-}$ obeying the following
anticommutation relations,%
\begin{equation}
\begin{tabular}{llll}
$\left\{ Q_{\alpha }^{+},Q_{\beta }^{+}\right\} $ & $=$ & $Z_{\left( \alpha
\beta \right) }^{++}$ & , \\
$\left\{ Q_{\alpha }^{+},\bar{Q}_{\dot{\beta}}^{-}\right\} $ & $=$ & $%
Z_{\alpha \dot{\beta}}^{0}$ & , \\
$\left\{ \bar{Q}_{\dot{\alpha}}^{-},\bar{Q}_{\dot{\beta}}^{-}\right\} $ & $=$
& $\bar{Z}_{\left( \dot{\alpha}\dot{\beta}\right) }^{--}$ & ,%
\end{tabular}
\label{an}
\end{equation}%
where the bosonic operators $Z_{\left( \alpha \beta \right) }^{++}$, $\bar{Z}%
_{\left( \dot{\alpha}\dot{\beta}\right) }^{--}$ are as before and where $%
Z_{\alpha \dot{\beta}}^{0}$ contains the usual energy momentum vector $%
P_{\mu }$ generating space time translations.

\subsection{More on central charges in \emph{8D} $\mathcal{N}=1$
supersymmetry}

The bosonic operators $Z_{\left( \alpha \beta \right) }^{++}$, $\bar{Z}%
_{\left( \dot{\alpha}\dot{\beta}\right) }^{--}$, $Z_{\alpha \dot{\beta}}^{0}$
capture several irreducible $SO\left( 1,7\right) $ space time
representations. To get their irreducible components, we use the
correspondence\footnote{%
viewed from \emph{4D}, this corresponds to $\mathcal{N}=4$ supersymmetry
with fermionic generators $Q_{\alpha }^{I}$ in the $\left( 2_{s},4\right) $
representation of $SO\left( 1,3\right) \times SU\left( 4\right) $.}
\begin{equation}
\begin{tabular}{llllllll}
$Q_{\alpha }^{+}$ & $\sim $ & $\left( 8_{s},+1\right) $ & , & $\bar{Q}_{\dot{%
\alpha}}^{-}$ & $\sim $ & $\left( 8_{c},-1\right) $ &
\end{tabular}%
\end{equation}%
and tensor product properties of the $SO\left( 1,7\right) \times U_{R}\left(
1\right) $ representations; in particular%
\begin{equation}
\begin{tabular}{llll}
$\left( 8_{s},+1\right) \times \left( 8_{s},+1\right) $ & $=$ & $\left(
1,+2\right) +\left( 28,+2\right) +\left( 35_{s},+2\right) $ & , \\
$\left( 8_{s},+1\right) \times \left( 8_{c},-1\right) $ & $=$ & $\left(
8_{v},0\right) +\left( 56_{v},0\right) $ & , \\
$\left( 8_{c},-1\right) \times \left( 8_{c},-1\right) $ & $=$ & $\left(
1,-2\right) +\left( 28,-2\right) +\left( 35_{c},-2\right) $ & .%
\end{tabular}
\label{8V}
\end{equation}%
The symmetry property of the anticommutators of eqs(\ref{an}) allows to read
the group theoretical structure of the $Z_{\left( \alpha \beta \right)
}^{++} $, $\bar{Z}_{\left( \dot{\alpha}\dot{\beta}\right) }^{--}$ and $%
Z_{\alpha \dot{\beta}}^{0}$; we have:
\begin{equation}
\begin{tabular}{llllll}
$Z_{\left( \alpha \beta \right) }^{++}$ & $\sim $ & $\left( 1,+2\right) $ & $%
\oplus $ & $\left( 35_{s},+2\right) $ & , \\
$Z_{\alpha \dot{\beta}}^{0}$ & $\sim $ & $\left( 8_{v},0\right) $ & $\oplus $
& $\left( 56_{v},0\right) $ & , \\
$\bar{Z}_{\left( \dot{\alpha}\dot{\beta}\right) }^{--}$ & $\sim $ & $\left(
1,-2\right) $ & $\oplus $ & $\left( 35_{c},-2\right) $ & .%
\end{tabular}
\label{35V}
\end{equation}%
Notice that the sub-index $i=v,s,c$ refer to the triality property of the $%
SO\left( 1,7\right) $ symmetry which have three kinds of fundamental
representations with same dimension. \newline
Moreover, using the $SO\left( 1,7\right) $ Dynkin labels $\left(
l_{1}l_{2}l_{3}l_{4}\right) $, the three eight dimensional basic
representations read as follows,%
\begin{equation}
\begin{tabular}{lll}
$8_{v}=\left( 1000\right) ,$ & $8_{s}=\left( 0001\right) ,$ & $8_{c}=\left(
0010\right) .$%
\end{tabular}%
\end{equation}%
With these basic representations, one can build the higher dimensional ones
by taking tensor products. For the example of the leading lower dimensional
representations, we have%
\begin{equation}
\begin{tabular}{llll}
$8_{i}\times 8_{i}$ & $=$ & $1+28+35_{i}$ & , \\
$8_{i}\times 8_{j}$ & $=$ & $8_{k}+56_{k}$ & ,%
\end{tabular}%
\end{equation}%
with $i,j,k$ cyclic and where%
\begin{equation}
\begin{tabular}{lll}
$35_{v}=\left( 2000\right) ,$ & $35_{s}=\left( 0002\right) ,$ & $%
35_{c}=\left( 0020\right) ,$ \\
$56_{v}=\left( 0011\right) ,$ & $56_{s}=\left( 1010\right) ,$ & $%
56_{c}=\left( 1001\right) .$%
\end{tabular}
\label{35S}
\end{equation}%
Notice also that besides the real energy momentum vector $P_{\mu }\sim 8_{v}$
and complex singlets $Z_{0}^{++}=Tr\left( Z_{\alpha \beta }^{++}\right) \sim
1$, the bosonic operators $Z_{\left( \alpha \beta \right) }^{++}$, $\bar{Z}%
_{\left( \dot{\alpha}\dot{\beta}\right) }^{--}$, $Z_{\alpha \dot{\beta}}^{0}$
capture moreover $SO\left( 1,7\right) $ higher dimensional representations
namely the $35_{s}$, $35_{c}$, $56_{v}$. \newline
In terms of $SO\left( 1,7\right) $ vector indices, these representations may
be decomposed by using \emph{antisymmetric} products of the $8\times 8$
Pauli-\textrm{Dirac} $\Gamma ^{\mu }$-\ matrices as follows%
\begin{equation}
\begin{tabular}{llll}
$Z_{\left( \alpha \beta \right) }^{++}$ & $=$ & $\delta _{\alpha \beta
}Z_{0}^{++}+\Gamma _{\left( \alpha \beta \right) }^{\mu \nu \rho \sigma }Z_{%
\left[ \mu \nu \rho \sigma \right] }^{++}$ & , \\
$\bar{Z}_{\left( \dot{\alpha}\dot{\beta}\right) }^{--}$ & $=$ & $\delta _{%
\dot{\alpha}\dot{\beta}}\bar{Z}_{0}^{--}+\Gamma _{\left( \alpha \beta
\right) }^{\mu \nu \rho \sigma }\tilde{Z}_{\left[ \mu \nu \rho \sigma \right]
}^{--}$ & , \\
$Z_{\alpha \dot{\beta}}^{0}$ & $=$ & $\Gamma _{\alpha \dot{\beta}}^{\mu
}Z_{\mu }^{0}+\Gamma _{\alpha \dot{\beta}}^{\mu \nu \rho }Z_{\left[ \mu \nu
\rho \right] }^{0}$ & ,%
\end{tabular}
\label{z}
\end{equation}%
where antisymmetrization with respect to space time indices is understood.
Notice that an antisymmetric rank 4-tensor type $Z_{\left[ \mu \nu \rho
\sigma \right] }$ has in general $\frac{8!}{4!\times 4!}$ degrees of
freedom; but the 4- forms $Z_{\left[ \mu \nu \rho \sigma \right] }^{++}$ and
$\tilde{Z}_{\left[ \mu \nu \rho \sigma \right] }^{--}$ involved in (\ref{z})
capture each $35$ degrees of freedom associated with the self dual and
anti-self dual antisymmetric 4-rank tensors in \emph{8D} space time,
\begin{equation}
\begin{tabular}{llll}
$Z_{\mu _{{\scriptsize 1}}\mu _{{\scriptsize 2}}\mu _{{\scriptsize 3}}\mu _{%
{\scriptsize 4}}}^{++}$ & $=$ & $\varepsilon _{\mu _{{\scriptsize 1}}....\mu
_{{\scriptsize 8}}}Z^{++\mu _{{\scriptsize 5}}\mu _{{\scriptsize 6}}\mu _{%
{\scriptsize 7}}\mu _{{\scriptsize 8}}}$ & , \\
$\tilde{Z}_{\mu _{{\scriptsize 1}}\mu _{{\scriptsize 2}}\mu _{{\scriptsize 3}%
}\mu _{{\scriptsize 4}}}^{--}$ & $=$ & $-\varepsilon _{\mu _{{\scriptsize 1}%
}....\mu _{{\scriptsize 8}}}\tilde{Z}^{--\mu _{{\scriptsize 5}}\mu _{%
{\scriptsize 6}}\mu _{{\scriptsize 7}}\mu _{{\scriptsize 8}}}$ & .%
\end{tabular}%
\end{equation}%
From this group theoretical analysis, it follows amongst others the two
following features:\newline
(\textbf{1}) the simplest form of the non chiral \emph{8D} $\mathcal{N}=1$
supersymmetric algebra reads as follows,%
\begin{equation}
\begin{tabular}{llll}
$\left\{ Q_{\alpha }^{+},Q_{\dot{\beta}}^{-}\right\} $ & $\sim $ & $\Gamma
_{\alpha \dot{\beta}}^{\mu }P_{\mu }$ & , \\
$\left\{ Q_{\alpha }^{+},Q_{\beta }^{+}\right\} $ & $\sim $ & $\left\{ Q_{%
\dot{\alpha}}^{-},Q_{\dot{\beta}}^{-}\right\} =0$ & ,%
\end{tabular}%
\end{equation}%
and corresponds to switching off the p-forms $Z_{0}^{++}$, $Z_{\mu }^{0}$, $%
Z_{\left[ \mu \nu \rho \right] }^{0}$ and $Z_{\left[ \mu \nu \rho \sigma %
\right] }^{\pm \pm },$\newline
(\textbf{2}) there are no $Z_{\left[ \mu \nu \right] }^{\pm \pm }$
components in eq(\ref{z}); this means that in non chiral \emph{8D} $\mathcal{%
N}=1$ supergravity we should have%
\begin{equation}
\begin{tabular}{llll}
$Z_{\left[ \mu \nu \right] }^{++}=0$ & , & $Z_{\left[ \mu \nu \right]
}^{--}=0$ & ,%
\end{tabular}
\label{c}
\end{equation}%
showing in turn that the supergravity multiplet has 1-form and 2-form gauge
fields; but no 3-form gauge field. \newline
Below, we switch on these charges and study extremal black attractors in non
chiral \emph{8D} $\mathcal{N}=1$ supergravity arising from superstring
compactifications.

\subsection{Central charges and branes}

From the above analysis, we learn that the bosonic $Z$- generators appearing
in the generalized supersymmetric algebra (\ref{an}) exhibit a set of
remarkable properties; in particular the three following ones:\newline
(\textbf{1}) to the bosonic operators $Z_{\mu _{1}...\mu _{p}}$, which are
charged under $SO\left( 1,7\right) \times U_{R}\left( 1\right) $, we
associate a space time p-form operator density
\begin{equation}
\begin{tabular}{llll}
$\mathcal{Z}_{p}$ & $=$ & $\frac{1}{p!}dx^{\mu _{1}}{\scriptsize \wedge
...\wedge }dx^{\mu _{p}}Z_{\mu _{1}...\mu _{p}}$ & ,%
\end{tabular}%
\end{equation}%
together with the charge,
\begin{equation}
\begin{tabular}{ll}
$\mathcal{J}_{p}=\int\nolimits_{M_{p}}\mathcal{Z}_{p}$ & ,%
\end{tabular}
\label{cc}
\end{equation}%
where $M_{p}$ is a p-dimensional space time submanifold which may be thought
of as the world volume of a p-brane. \newline
(\textbf{2}) The $Z_{\mu _{1}...\mu _{p}}$ operators have an interpretation
in terms of fluxes of gauge fields in non chiral \emph{8D} supergravity. By
using the usual relations $m=\frac{1}{4\pi }\int\nolimits_{S^{2}}\mathcal{F}%
_{2}$ and $e=\frac{1}{4\pi }\int\nolimits_{S^{2}}\mathcal{\tilde{F}}_{2}$
giving the magnetic and electric charges of particles coupled to \emph{4D}\
Maxwell gauge fields and thinking about the $\mathcal{Z}_{p}$'s in the same
manner, we end with the following relations
\begin{equation}
\begin{tabular}{llllll}
$\mathcal{Z}_{0}\sim \int_{S^{2}}\mathcal{F}_{2}$ & , & $\mathcal{Z}_{1}\sim
\int_{S^{2}}\mathcal{F}_{3}$ & , & $\mathcal{Z}_{2}\sim \int_{S^{2}}\mathcal{%
F}_{4}$ &
\end{tabular}%
\end{equation}%
as well as their duals. In these relations, the $\mathcal{F}_{p}$'s stand
for the gauge invariant p-forms,%
\begin{equation}
\begin{tabular}{llll}
$\mathcal{F}_{2}=d\mathcal{A}_{1},$ & $\mathcal{F}_{3}=d\mathcal{A}_{2},$ & $%
\mathcal{F}_{4}=d\mathcal{A}_{3},$ &
\end{tabular}%
\end{equation}%
with Hodge duals%
\begin{equation}
\begin{tabular}{llll}
$\mathcal{\tilde{F}}_{4}=\left( ^{\star }\mathcal{F}_{4}\right) ,$ & $%
\mathcal{\tilde{F}}_{5}=\left( ^{\star }\mathcal{F}_{3}\right) ,$ & $%
\mathcal{\tilde{F}}_{6}=\left( ^{\star }\mathcal{F}_{2}\right) ,$ &
\end{tabular}%
\end{equation}%
from which we learn
\begin{equation}
\begin{tabular}{ll}
$\mathcal{J}_{p}=\int_{M_{p}\times S^{2}}\mathcal{F}_{p+2}$ & ,%
\end{tabular}
\label{ccc}
\end{equation}%
teaching us that the $\mathcal{Z}_{p}$'s describe precisely charges of
p-branes that couple to the \emph{8D} supergravity $\left( p+1\right) $-
form gauge fields $\mathcal{A}_{p+1}$ with the field strengths $\mathcal{F}%
_{p+2}$ and their magnetic duals $\mathcal{\tilde{F}}_{6-p}$. \newline
(\textbf{3}) Using the relation (\ref{c}) and eqs(\ref{cc}-\ref{ccc}) it
follows that $\mathcal{J}_{2}=0$ and
\begin{equation}
\begin{tabular}{lllll}
$\int_{M_{2}\times S^{2}}\mathcal{F}_{4}$ & $=$ & $\int_{\left( \partial
M_{2}\right) \times S^{2}}\mathcal{C}_{3}$ & $=0$ & ,%
\end{tabular}%
\end{equation}%
showing that, in non chiral $\mathcal{N}=1$\ supergravity, there is no
magnetic nor electric charges associated with the dyonic 4-form gauge
invariant field strength $\mathcal{F}_{4}=d\mathcal{C}_{3}$. In other words
there is no D2- brane in the type IIA set up of non chiral \emph{8D} $%
\mathcal{N}=1$\ supergravity.

\section{Fluxes of black attractors in \emph{8D}}

In this section we study the non chiral \emph{8D} $\mathcal{N}=1$
supergravity arising from low energy compactifications of \emph{10D}
superstring that preserve sixteen supersymmetric charges. \newline
We first study the case of \emph{8D} $\mathcal{N}\emph{=1}$ supergravity
embedded in heterotic string on $T^{2}$ with moduli space
\begin{equation}
\begin{tabular}{lllll}
$\boldsymbol{M}_{8D\text{-}Het/T^{2}}^{N=1}$ & $=\frac{SO\left( 2,r+2\right)
}{SO\left( 2\right) \times SO\left( r+2\right) }\times SO\left( 1,1\right) $
& , & $r\geq 0$ & .%
\end{tabular}%
\end{equation}%
Then we develop the dual type IIA superstring on a compact real surface $%
\Sigma _{2}^{\left( r\right) }$. In this case, we will focus on the class of
real surfaces given by the following union of irreducible 2-cycles
(2-spheres) $C_{I}$
\begin{equation}
\begin{tabular}{llll}
$\Sigma _{2}^{\left( r\right) }$ & $=$ & $C_{0}\cup \left(
\dbigcup\limits_{I=1}^{r-1}C_{I}\right) $ &
\end{tabular}
\label{a}
\end{equation}%
with intersection matrix
\begin{equation}
\begin{tabular}{llll}
$C_{I}.C_{I}$ & $=$ & $-K_{IJ}$ &
\end{tabular}%
\end{equation}%
coinciding with the Cartan matrix of the of simply laced ADE Lie algebras.
The moduli space of this theory is
\begin{equation}
\begin{tabular}{lllll}
$\boldsymbol{M}_{8D-IIA/\Sigma }^{N=1}$ & $=\frac{SO\left( 2,r+1\right) }{%
SO\left( 2\right) \times SO\left( r+1\right) }\times SO\left( 1,1\right) $ &
, & $r\geq 0$ & .%
\end{tabular}%
\end{equation}%
The simplest surface $\Sigma _{2}^{\left( r\right) }$ corresponds obviously
to taking $r=0$ and its singular limit given by $vol\left( C_{0}\right)
\rightarrow 0$ should be associated with a non abelian $SU\left( 2\right) $
gauge symmetry. The similarity between $\boldsymbol{M}_{8D\text{-}%
Het/T^{2}}^{N=1}$ and $\boldsymbol{M}_{8D-IIA/\Sigma }^{N=1}$ shows
precisely the duality between the two constructions; for details see \textrm{%
\cite{100}}.

\subsection{Heterotic string on T$^{2}$}

First recall that the massless bosonic fields of the \emph{10D} heterotic
string belong to two representations of the \emph{10D} supersymmetric
algebras; these are $\mathcal{G}_{MN}^{{\small 10D}},$ $\mathcal{B}_{MN}^{%
{\small 10D}},$ $\Phi _{dil}^{{\small 10D}}$ of the supergravity multiplet
and the typical gauge fields $\mathcal{A}_{M}^{I}$ belonging to the Yang
Mills multiplets. As we are interested in this study into black attractor
solutions, we will restrict below to the abelian sector and think about $%
\mathcal{A}_{M}^{I}$ as Maxwell gauge fields associated with the Cartan
subalgebra of a given rank $r$ gauge group; i.e: $I=1,...,r.$\newline
Under compactification of these bosonic fields on the two torus T$^{2}$, we
get the following $8D$ ones%
\begin{equation}
\begin{tabular}{llll}
$\mathcal{G}_{\mu \nu },$ & $\mathcal{B}_{\mu \nu },$ & $\sigma ,$ & $%
\mathcal{A}_{\mu }^{I},$%
\end{tabular}%
\end{equation}%
together with the four $8D$\ gauge fields
\begin{equation}
\begin{tabular}{ll}
$\mathcal{G}_{\mu }^{i}$, & $\mathcal{B}_{\mu }^{i}$,%
\end{tabular}%
\end{equation}%
as well as the $\left( 4+2r\right) $ scalars
\begin{equation}
\begin{tabular}{lll}
$\mathcal{G}^{\left( ij\right) }$, & $\mathcal{B}^{\left[ ij\right]
}=\varepsilon ^{ij}B,$ & $\mathcal{A}^{iI}$.%
\end{tabular}%
\end{equation}%
These fields combine into two \emph{8D} $\mathcal{N}=1$ supermultiplets
namely:

\begin{itemize}
\item the \emph{8D}\ gravity multiplet with bosonic content%
\begin{equation}
\begin{tabular}{llll}
$\mathcal{G}_{\mu \nu },$ & $\mathcal{B}_{\mu \nu },$ & $\mathcal{C}_{\mu
}^{i},$ & $\sigma $%
\end{tabular}%
\end{equation}%
containing the \emph{8D} graviton $\mathcal{G}_{\mu \nu }$, the $\mathcal{B}%
_{\mu \nu }$ antisymmetric field, two gauge fields $\mathcal{C}_{\mu
}^{i}=\left( \mathcal{C}_{\mu }^{1},\mathcal{C}_{\mu }^{2}\right) $\
transforming as a real 2-vector under $SO\left( 2\right) $ R-symmetry; and
the \emph{8D}\ dilaton $\sigma $. \newline
The total number of the degrees of freedom of this gravity multiplet is
\emph{48+48}; the other \emph{48} superpartners come from the gravitino $%
\Psi _{\mu \alpha }$ and a photino $\chi _{\alpha }$ carrying respectively
\emph{40} and \emph{8} fermionic degrees of freedom.

\item the \emph{8D} Maxwell multiplets whose bosonic fields are given by%
\begin{equation}
\begin{tabular}{llll}
$\mathcal{A}_{\mu }^{i},$ & $\phi ^{ij},$ & $\mathcal{A}_{\mu }^{I},$ & $%
\phi ^{iI}$.%
\end{tabular}
\label{gau}
\end{equation}%
These \emph{8D} $\mathcal{N}=1$ supermultiplets contain $\left( 2+r\right) $
Maxwell gauge fields ($\mathcal{A}_{\mu }^{i},\mathcal{A}_{\mu }^{I}$) which
we denote collectively as $\mathcal{A}_{\mu }^{a}$ with $a=1,...r+2$; and $%
2\left( r+2\right) $ real scalars $\phi ^{ia}\equiv \left( \phi ^{ij},\phi
^{iI}\right) $. Together with these bosons, we also have $r+2$ gauginos $%
\lambda _{\alpha }^{a}$ given by pseudo-Majorana spinors in \emph{8D}.
\end{itemize}

\ \ \ \ \ \newline
The moduli space of this \emph{8D} $\mathcal{N}=1$ supergravity that is
embedded heterotic superstring on T$^{2}$ reads as follows%
\begin{equation}
\begin{tabular}{llll}
$\boldsymbol{M}_{8D\text{-}Het/T^{2}}^{N=1}$ & $=$ & $\frac{SO\left(
2,r+2\right) }{SO\left( 2\right) \times SO\left( r+2\right) }\times SO\left(
1,1\right) $ &
\end{tabular}
\label{het}
\end{equation}%
where the extra factor $SO\left( 1,1\right) $ refers to the dilaton $\sigma $
and $\frac{SO\left( 2,r+2\right) }{SO\left( 2\right) \times SO\left(
r+2\right) }$ for $\phi ^{ia}$. This real space has $\left( 2r+5\right) $
dimensions; it reduces for the particular case $r=0$, to the five
dimensional one%
\begin{equation}
\begin{tabular}{l}
$\frac{SO\left( 2,2\right) }{SO\left( 2\right) \times SO\left( 2\right) }%
\times SO\left( 1,1\right) $%
\end{tabular}
\label{H}
\end{equation}%
In addition to the dilaton, the four scalars $\phi ^{ij}$ have geometric and
stringy interpretations; three of them are given by the Kahler and complex
structure of the 2-torus; the fourth is given by the value of the B$_{NS}$
field on $T^{2}$.\newline
The field strengths associated with the various gauge fields of the \emph{8D}
supergravity are given by the gauge invariant forms
\begin{equation}
\begin{tabular}{llllll}
$\mathcal{F}_{2}^{i}=d\mathcal{C}_{1}^{i}$ & , & $\mathcal{F}_{2}^{a}=d%
\mathcal{A}_{1}^{a}$ & , & $\mathcal{F}_{3}=d\mathcal{B}_{2}$ & ,%
\end{tabular}
\label{u}
\end{equation}%
For later use, we give below the magnetic and electric charges associated
with these field strengths as well as the brane interpretation; more details
will be given when we consider the type IIA dual derivation. We have:

\begin{itemize}
\item a black hole and its 4-brane dual associated with the two graviphotons
$\mathcal{C}_{\mu }^{i}$ with magnetic and electric charges as follows
\begin{equation}
\begin{tabular}{lllll}
$g^{i}=\int_{S^{2}}\mathcal{F}_{2}^{i}$ & , & $e_{i}=\int_{\tilde{S}^{6}}%
\mathcal{\tilde{F}}_{6|i}$ &  &
\end{tabular}%
\end{equation}

\item a black hole and its 4-brane dual associated with the $\left(
r+2\right) $ Maxwell fields $\mathcal{A}_{\mu }^{a}$; their magnetic and
electric charges are given by
\begin{equation}
\begin{tabular}{lllll}
$p^{a}=\int_{S^{2}}\mathcal{F}_{2}^{a}$ & , & $q_{a}=\int_{\tilde{S}^{6}}%
\mathcal{\tilde{F}}_{6|a}$ &  &
\end{tabular}%
\end{equation}%
these two kinds of magnetic and electric charges of the black hole/black
4-brane combine into $SO\left( 2,r+2\right) $ vector charges as given below
\begin{equation}
\begin{tabular}{llll}
$P^{\Lambda }=\left( g^{i},p^{a}\right) $ & , & $Q_{\Lambda }=\left(
e_{i},q_{a}\right) $ &
\end{tabular}
\label{PQ}
\end{equation}

\item a black string and its 3-brane dual associated with the B$_{\mu \nu }$%
-field; the corresponding charges are given by%
\begin{equation}
\begin{tabular}{llll}
$p^{0}=\int_{S^{3}}\mathcal{F}_{3}$ & , & $q_{0}=\int_{\tilde{S}^{5}}%
\mathcal{\tilde{F}}_{5}$ &
\end{tabular}%
\end{equation}
\end{itemize}

\ \ \ \newline
All these electric and magnetic charges are linked by the usual Dirac
quantization relation; they determine the effective potential
\begin{equation}
\mathcal{V}_{eff}^{het}=\mathcal{V}_{eff}^{het}\left(
P,Q;p^{0},q_{0},...\right)
\end{equation}%
of the \emph{8D} black attractors to be considered later.\newline
Notice that one of the remarkable features of this analysis is the absence
of the 4-form field strengths $\mathcal{F}_{4}$, $\mathcal{\tilde{F}}_{4}$
as predicted from the group theory view. In what follows, we explore this
issue by studying the type IIA dual compactification down to \emph{8D}.

\subsection{Black attractors in type IIA on $\Sigma _{2}^{\left( r\right) }$}

In this subsection, we study the embedding of non chiral \emph{8D} $\mathcal{%
N}=1$ supergravity in type IIA superstring on $\Sigma _{2}^{\left( r\right)
} $. To that purpose, we first study the compactification of type IIA on the
2-sphere $\mathcal{S}^{2}$ corresponding to $\Sigma _{2}^{\left( 0\right) }$%
. Then, we extend the analysis to the surface $\Sigma _{2}^{\left( r\right)
} $ given by eq(\ref{a}). \newline
In type ten dimensional type IIA superstring with \emph{32} supercharges,
the massless bosonic particles of the perturbative spectrum, describing the
low energy \emph{10D }type IIA supergravity, is given by
\begin{equation}
\begin{tabular}{lllllll}
bosons & $:$ & $\mathcal{G}_{MN}$, & $\mathcal{B}_{MN},$ & $\Phi _{dil},$ & $%
\mathcal{A}_{M}$, & $\mathcal{C}_{MNK},$%
\end{tabular}
\label{10}
\end{equation}%
with indices $M,N,K=0,\ldots ,9$ transforming as $SO\left( 1,9\right) $
vectors. \newline
These fields capture \emph{128} on shell degrees of freedom partitioned as
follows%
\begin{equation}
\begin{tabular}{ll}
$128=64+64=\left( 35+28+1\right) +\left( 8+56\right) $ & ,%
\end{tabular}%
\end{equation}%
with the first \emph{64} coming from NS-NS sector ($\mathcal{G}_{MN},$ $%
\mathcal{B}_{MN},$ $\Phi _{dil}$) and the other \emph{64} from RR-sector;
i.e ($\mathcal{A}_{M}$, $\mathcal{C}_{MNK}$).\newline
We also have a non perturbative sector with p-branes namely%
\begin{equation}
\begin{tabular}{lllllll}
F1 string, & NS 5-brane & ; & D0, & D2, & D4, & D6.%
\end{tabular}%
\end{equation}%
Some of these branes are the source of the gauge fields involved in the
Maxwell sector of non chiral \emph{8D} $\mathcal{N}=1$ supergravity. The
fields are mainly similar to those given by eq(\ref{gau}); but here they
should be thought of as the gauge fields associated with a D2- brane wrapped
the irreducible 2-cycles of $\Sigma _{2}^{\left( r\right) }$.

\subsubsection{Compactification of type IIA on \emph{S}$^{2}$}

After the space time compactification $R^{1,9}\rightarrow R^{1,7}\times
S^{2} $ where the local coordinates $\left( x^{0},...,x^{9}\right) $ get
split as $\left( x^{0},...,x^{7}\right) $ and $y=\left( z,\bar{z}\right) $
with $z=x^{8}+ix^{9}$ parameterizing the 2-sphere, the bosonic fields of the
spectrum (\ref{10}) reduces to:%
\begin{equation}
\begin{tabular}{llllllll}
$\left\{
\begin{array}{c}
\mathcal{G}_{\mu \nu } \\
\phi ^{1}%
\end{array}%
\right. ,$ & $\left\{
\begin{array}{c}
\mathcal{B}_{\mu \nu } \\
\phi ^{2}%
\end{array}%
\right. ,$ & $\sigma $ & $;$ & $\mathcal{A}_{\mu },$ & $\mathcal{C}_{\mu \nu
\rho },$ & $\mathcal{C}_{\mu },$ &
\end{tabular}%
\end{equation}%
with
\begin{equation}
\begin{tabular}{llllll}
$\phi ^{1}=\mathcal{G}_{z\bar{z}}$ & $,$ & $\phi ^{2}=\mathcal{B}_{z\bar{z}}$
& , & $\mathcal{C}_{\mu }=\mathcal{C}_{\mu }^{z\bar{z}}$ &
\end{tabular}%
\end{equation}%
respectively describing the Kahler modulus of the 2-sphere, the B$_{NS}$
field on \emph{S}$^{2}$ and the gauge particle associated with a D2-brane
wrapping $S^{2}$. \newline
This field spectrum has \emph{70} bosonic degrees of freedom; but only \emph{%
48} of them combine with the \emph{8D} gravitino $\Psi _{\mu }=\left( \Psi
_{\mu }^{\alpha },\bar{\Psi}_{\mu \dot{\alpha}}\right) $ and the
graviphotino $\chi =\left( \chi ^{\alpha },\bar{\chi}_{\dot{\alpha}}\right) $
to form the non chiral $\mathcal{N}=1$ supergravity multiplet%
\begin{equation}
\begin{tabular}{llll}
$\mathcal{G}_{\mu \nu },\mathcal{B}_{\mu \nu },\sigma ,\mathcal{A}_{\mu
}^{i} $ & $;$ & $\Psi _{\mu },\chi $ &
\end{tabular}
\label{gr}
\end{equation}%
where $\mathcal{A}_{\mu }^{i}$ stands for the $SO\left( 2\right) $ doublet $%
\left( \mathcal{A}_{\mu },\mathcal{C}_{\mu }\right) $. The on shell degrees
of freedom are partitioned as $48_{\text{bose}}=\left( 20+15+1\right)
+2\times 6$ and $48_{\text{fermi}}=40+8$. We also have the following branes,%
\begin{equation}
\begin{tabular}{lllllll}
{\small F1 string,} & $\left( \text{{\small NS 5-brane}}{\small /}\text{%
{\small S}}^{{\small 2}}\right) $ & ; & {\small D0,} & $\left( \text{{\small %
D6/S}}^{{\small 2}}\right) ,$ & $\left( \text{{\small D2/S}}^{{\small 2}%
}\right) $, & $\left( \text{{\small D4/S}}^{{\small 2}}\right) .$%
\end{tabular}%
\end{equation}%
satisfying the usual \emph{8D} electric magnetic duality relation between
electric q-brane and its magnetic p-brane dual with the integers p and q
constrained as $p+q=4$.

\subsubsection{Compactification of type IIA on $\Sigma _{2}^{\left( r\right)
}$}

The compactification of field content of the type IIA superstring on $\Sigma
_{2}^{\left( r\right) }$ extends the case of the 2-sphere; it leads to the
following:\newline
(\textbf{1}) a gravity multiplet; the same as in eq(\ref{gr})\newline
(\textbf{2}) r Maxwell multiplets given by%
\begin{equation}
\begin{tabular}{lll}
$\mathcal{A}_{\mu }^{a},$ & $\phi ^{ia},$ & $a=1,...,r$%
\end{tabular}%
\end{equation}%
where the $\phi ^{1a}$'s are the Kahler parameters of the irreducible
2-cycles C$_{a}$ (2-spheres S$_{a}^{2}$) involved in $\Sigma _{2}^{\left(
r\right) }$ and the $\phi ^{1a}$'s stand for the values of the B$_{NS}$
fields on these S$_{a}^{2}$'s. \newline
The gauge fields $\mathcal{A}_{\mu }^{a}$ are associated with the wrapping
of D2-brane on the S$_{a}^{2}$'s of the compact surface $\Sigma _{2}^{\left(
r\right) }$; i.e:%
\begin{equation}
\begin{tabular}{llll}
$\mathcal{A}_{\mu }^{a}$ & : & $\left( \text{D2/}S_{a}^{2}\right) $ &
\end{tabular}%
\end{equation}%
The moduli space $\boldsymbol{M}_{8D\text{-}IIA/\Sigma _{2}^{r}}^{N=1}$ of
this \emph{8D} $\mathcal{N}=1$ supergravity is parameterized by the $\left(
2r+1\right) $ scalars namely the dilaton $\sigma $ and the $SO\left(
2\right) $ doublets $\phi ^{ia}$; it is given by
\begin{equation}
\begin{tabular}{ll}
&  \\
$\boldsymbol{M}_{8D\text{-}IIA/\Sigma _{2}^{r}}^{N=1}$ & $\boldsymbol{=}%
\frac{SO\left( 2,r\right) }{SO\left( 2\right) \times SO\left( r\right) }%
{\small \times SO}\left( {\small 1,1}\right) $ \\
&
\end{tabular}
\label{pm}
\end{equation}%
This space is comparable to $\boldsymbol{M}_{8D\text{-het}/T^{2}}^{N=1}$ eq(%
\ref{het}); this is due to the string-string duality between the heterotic
on T$^{2}$ and type IIA superstring on $\Sigma _{2}^{\left( r\right) }$ that
follows from the well known duality relation in \emph{6D }space time,
\begin{equation}
\begin{tabular}{llll}
{\small Het/T}$^{4}$ & $\leftrightarrow $ & {\small Type IIA/K3} &
\end{tabular}%
\end{equation}%
The field strengths $\mathcal{F}_{p+1}=d\mathcal{A}_{p}$, associated with
the various gauge fields of the \emph{8D} $\mathcal{N}=1$ supergravity, are
given by
\begin{equation}
\begin{tabular}{lllll}
$\mathcal{F}_{2}^{i}=d\mathcal{A}^{i}$ & , & $\mathcal{F}_{2}^{a}=d\mathcal{A%
}^{a}$ & , & $\mathcal{F}_{3}=d\mathcal{B}_{2}$%
\end{tabular}%
\end{equation}%
These gauge invariant fields transform under the $SO\left( 2\right) \times
SO\left( r\right) $ group as follows%
\begin{equation}
\begin{tabular}{l|l|l|l}
& $\mathcal{F}_{2}^{i}$ & $\mathcal{F}_{2}^{a}$ & $\mathcal{F}_{3}$ \\ \hline
${\small SO}\left( {\small 2}\right) \times {\small SO}\left( {\small r}%
\right) $ & $\left( 2,1\right) $ & $\left( 1,r\right) $ & $\left( 1,1\right)
$%
\end{tabular}%
\end{equation}%
The corresponding magnetic and electric charges are as follows:\newline
(\textbf{a}) the string and its \textrm{dual 3-brane}%
\begin{equation}
\begin{tabular}{llll}
$p^{0}=\int_{S^{3}}\mathcal{F}_{3}$ & , & $q_{0}=\int_{\tilde{S}^{5}}%
\mathcal{\tilde{F}}_{5}$ &
\end{tabular}
\label{pq}
\end{equation}%
the string is magnetically charged while the 3-brane is electrically charged.%
\newline
(\textbf{b}) the black hole magnetic charges $\left( {\small g}^{i},{\small p%
}^{a}\right) $ and the electric duals $\left( e_{i},q_{a}\right) $ given by%
\begin{equation}
\begin{tabular}{llll}
${\small g}^{i}=\int_{S^{2}}\mathcal{F}_{2}^{i}$ & , & $e_{i}=\int_{\tilde{S}%
^{6}}\mathcal{\tilde{F}}_{6|i}$ & , \\
${\small p}^{a}=\int_{S_{a}^{2}}\mathcal{F}_{2}$ & , & $q_{a}=\int_{\tilde{S}%
_{a}^{6}}\mathcal{\tilde{F}}_{6|a}$ & .%
\end{tabular}
\label{bar}
\end{equation}%
These are respectively $SO\left( 2\right) $ and $SO\left( r\right) $ vectors.%
\newline
We end this section by noting that one may write down the lagrangian
densities of the various gauge fields. For the bosonic sector we have, in
addition to the Einstein-Hilbert term\textrm{\ }$\frac{1}{16\pi G_{8}}\int_{%
\mathcal{M}_{8}}\sqrt{-\mathcal{G}}\mathcal{R}_{8D}$, two other
contributions; the first one is given by the 1-form gauge fields\textrm{\ }$%
\mathcal{A}_{\mu }^{\Lambda }=\left( \mathcal{A}_{\mu }^{i},\mathcal{A}_{\mu
}^{a}\right) $%
\begin{equation}
\begin{tabular}{ll}
$\mathcal{L}_{\text{{\scriptsize 1-form}}}=\frac{1}{16\pi G_{8}}\int_{%
\mathcal{M}_{8}}\sqrt{-\mathcal{G}}$ $\left[ \mathcal{N}_{ab}\text{ }%
\mathcal{F}_{\mu \nu }^{a}\mathcal{F}^{\mu \nu b}+\mathcal{N}_{ij}\text{ }%
\mathcal{F}_{\mu \nu }^{i}\mathcal{F}^{\mu \nu j}\right] $ & ,%
\end{tabular}%
\ \
\end{equation}%
where the field metric $\mathcal{N}_{\Lambda \Gamma }=\mathcal{N}_{\Lambda
\Gamma }\left( \phi ,\sigma \right) $ reads as $e^{2\sigma }L_{\Lambda
}^{i}\left( \phi \right) \delta _{ij}L_{\Gamma }^{j}\left( \phi \right) $ $-$
$e^{2\sigma }L_{\Lambda }^{a}\left( \phi \right) \eta _{ab}L_{\Gamma
}^{b}\left( \phi \right) $ with the field matrix $L_{\Lambda \Upsilon }$
parameterizing the $SO\left( 2,N\right) $ group [see eqs(\ref{sa}-\ref{so})]
and $\phi \equiv \left\{ \phi ^{ia}\right\} $ being the free moduli that
parameterize the moduli space $\frac{SO\left( 2,N\right) }{SO\left( 2\right)
\times SO\left( N\right) }$. The second contribution comes from the $%
\mathcal{B}_{2}$ gauge field; it reads as follows%
\begin{equation}
\begin{tabular}{ll}
$\mathcal{L}_{\text{{\scriptsize 2-form}}}=\frac{1}{16\pi G_{8}}\int_{%
\mathcal{M}_{8}}\sqrt{-\mathcal{G}}$ $\mathcal{N}\left( \sigma \right)
\mathcal{F}_{\mu \nu \rho }\mathcal{F}^{\mu \nu \rho }$ & ,%
\end{tabular}
\label{3}
\end{equation}%
where now $\mathcal{N}\left( \sigma \right) $ is the metric associated with
the $SO\left( 1,1\right) $ factor parameterized by the dilaton $\sigma $
with no dependence in the $\phi ^{ia}$s.

\section{Attractor equations and solutions}

We first describe the effective potential of the black branes in non chiral
\emph{8D} $\mathcal{N}=1$ supergravity. Then, we study the attractor eqs and
their solutions.

\subsection{Effective potential}

The effective potential $\mathcal{V}_{eff}$ of the black attractors has
contributions coming from the various gauge fields of the non chiral \emph{8D%
} $\mathcal{N}=1$ supergravity. As there is no contribution coming from the
D2-brane we have:%
\begin{equation}
\begin{tabular}{lll}
$\mathcal{V}_{eff}=$ & $\left( \mathcal{V}_{BH}+\mathcal{V}_{4B}\right)
+\left( \mathcal{V}_{string}+\mathcal{V}_{3B}\right) $ & ,%
\end{tabular}%
\end{equation}%
with: (\textbf{a}) $\mathcal{V}_{BH}$ is the effective potential of the
\emph{8D }black hole associated with the charges of the two graviphotons $%
\mathcal{A}_{\mu }^{i}$ and the Maxwell gauge fields $\mathcal{A}_{\mu }^{a}$%
. It is given by the usual Weinhold relation whose expression, in the flat
coordinate frame, reads as:%
\begin{equation}
\begin{tabular}{lll}
$\mathcal{V}_{BH}=$ & $\dsum\limits_{i,j=1}^{2}\delta
_{ij}X^{i}X^{j}+\dsum\limits_{a,b=1}^{N}\delta _{ab}Y^{a}Y^{b}$ &
\end{tabular}%
\end{equation}%
In this relation, $X^{i}$ and $Y^{a}$, which respectively transform as $%
SO\left( 2\right) $ and $SO\left( N\right) $ vectors, are the dressed
central charges related to the magnetic bare $P^{\Lambda }=\left(
g^{i},p^{a}\right) $ of eqs(\ref{bar}) like,%
\begin{equation}
\begin{tabular}{llll}
$X^{i}=e^{\sigma }L_{\Lambda }^{i}P^{\Lambda }$ & , & $Y^{a}=e^{\sigma
}L_{\Lambda }^{a}P^{\Lambda }$ & .%
\end{tabular}
\label{yy}
\end{equation}%
In these relations, $e^{\sigma }$ and $L_{\Lambda \Upsilon }$ parameterize
respectively the $SO\left( 1,1\right) $ and $SO\left( 2,N\right) $ group
factors of the moduli space of the non chiral \emph{8D} $\mathcal{N}=1$
supergravity. Notice that the $L_{\Lambda \Upsilon }$ matrix is a real $%
\left( 2+N\right) \times \left( 2+N\right) $ matrix
\begin{equation}
\begin{tabular}{ll}
$L_{\Lambda }^{\Gamma }=\left(
\begin{array}{cc}
L_{i}^{j} & L_{i}^{b} \\
L_{a}^{j} & L_{a}^{b}%
\end{array}%
\right) $ & ,%
\end{tabular}
\label{sa}
\end{equation}%
satisfying the usual orthogonality relation $L^{T}\eta L=\eta $ which
explicitly reads like
\begin{equation}
\begin{tabular}{lll}
$L_{\Gamma \Lambda }\eta ^{\Gamma \Sigma }L_{\Sigma \Upsilon }=\eta
_{\Lambda \Upsilon }$ & , &
\end{tabular}
\label{so}
\end{equation}%
with $\eta _{\Lambda \Upsilon }=diag\left( +,+,-,\cdots ,-\right) $. A
priori $L_{\Lambda \Gamma }$ has $\left( 2+N\right) ^{2}$ parameters; but
this relation may be viewed as a constraint relation that reduces this
number down to $\frac{1}{2}\left( N^{2}+3N+2\right) $. Furthermore
subtracting the $\frac{1}{2}\left( N^{2}-N+2\right) $ gauge degrees of
freedom captured by the $SO\left( 2\right) \times SO\left( N\right) $
symmetry of the moduli space, we end with $2N$ moduli parameterizing $\frac{%
SO\left( 2,N\right) }{SO\left( 2\right) \times SO\left( N\right) }$.\newline
(\textbf{b}) $\mathcal{V}_{4B}$ is the effective potential associated with
the black 4D- branes dual the black holes,
\begin{equation}
\begin{tabular}{lll}
$\mathcal{V}_{4B}=$ & $\dsum\limits_{i,j=1}^{2}\delta ^{ij}\tilde{X}_{i}%
\tilde{X}_{j}+\dsum\limits_{a,b=1}^{N}\delta ^{ab}\tilde{Y}_{a}\tilde{Y}_{b}$
&
\end{tabular}%
\end{equation}%
The $\tilde{X}_{i}$ and $\tilde{Y}_{a}$ are dressed central charges related
to the electric $Q_{\Lambda }=\left( e_{i},q_{a}\right) $ as follows%
\begin{equation}
\begin{tabular}{llll}
$\tilde{X}_{i}=Q_{\Lambda }\left( L^{-1}\right) _{i}^{\Lambda }e^{-\sigma }$
& , & $\tilde{Y}_{a}=Q_{\Lambda }\left( L^{-1}\right) _{a}^{\Lambda
}e^{-\sigma }$ &
\end{tabular}
\label{xx}
\end{equation}%
(\textbf{c}) the term $\left( \mathcal{V}_{string}+\mathcal{V}_{3B}\right) $
is the effective potential of the black string and its 3- brane (NS 5-brane/S%
$^{2}$) dual; it is given by%
\begin{equation}
\begin{tabular}{llll}
$\mathcal{V}_{string}=e^{4\sigma }p_{0}^{2}$ & , & $\mathcal{V}%
_{3B}=e^{-4\sigma }q_{0}^{2}$ &
\end{tabular}%
\end{equation}%
where the magnetic charge p$_{0}$ and the electric q$_{0}$ one are as in eq(%
\ref{pq}).\newline
Adding all terms, we get the total effective potential of the black
attractors in non chiral \emph{8D} $\mathcal{N}=1$ supergravity%
\begin{equation}
\begin{tabular}{ll}
$\mathcal{V}_{eff}=$ & $+\dsum\limits_{i,j=1}^{2}\left( X^{i}\delta
_{ij}X^{j}+\tilde{X}_{i}\delta ^{ij}\tilde{X}_{j}\right) +\left( e^{4\sigma
}p_{0}^{2}+e^{-4\sigma }q_{0}^{2}\right) $ \\
& $+\dsum\limits_{a,b=1}^{N}\left( Y^{a}\delta _{ab}Y^{b}+\tilde{Y}%
_{a}\delta ^{ab}\tilde{Y}_{b}\right) $%
\end{tabular}
\label{VEF}
\end{equation}%
It is manifestly invariant under the $SO\left( 2\right) \times SO\left(
N\right) $ symmetry of the moduli space.\newline
Substituting the dressed central charges $X^{i},$ $\tilde{X}_{i},$ $Y^{a}$, $%
\tilde{Y}_{a}$ by their explicit expressions in terms of the field moduli,
we end with a function depending on the electric and magnetic charges as
well as on the scalars $\sigma $ and $L_{\Lambda M}$,%
\begin{equation}
\begin{tabular}{lll}
$\mathcal{V}_{eff}=$ & $\mathcal{V}_{eff}\left( \sigma ,L_{\Lambda
M};P^{\Lambda },Q_{\Lambda };p_{0},q_{0}\right) $ & .%
\end{tabular}%
\end{equation}%
More explicitly, we have%
\begin{equation}
\begin{tabular}{ll}
$\mathcal{V}_{eff}=$ & $+\dsum\limits_{i,j=1}^{2}\left( e^{2\sigma
}P^{\Lambda }L_{\Lambda }^{i}\delta _{ij}L_{\Upsilon }^{j}P^{\Upsilon
}+e^{-2\sigma }Q_{\Lambda }\left( L^{-1}\right) _{i}^{\Lambda }\delta
^{ij}\left( L^{-1}\right) _{j}^{\Upsilon }Q_{\Upsilon }\right) $ \\
& $+\dsum\limits_{a,b=1}^{N}\left( e^{2\sigma }P^{\Lambda }L_{\Lambda
}^{a}\delta _{ab}L_{\Upsilon }^{b}P^{\Upsilon }+e^{-2\sigma }Q_{\Lambda
}\left( L^{-1}\right) _{a}^{\Lambda }\delta ^{ab}\left( L^{-1}\right)
_{b}^{\Upsilon }Q_{\Upsilon }\right) $ \\
& $+\left( e^{4\sigma }p_{0}^{2}+e^{-4\sigma }q_{0}^{2}\right) $%
\end{tabular}%
\end{equation}%
with $L_{\Lambda \Upsilon }$ belonging to $SO\left( 2,N\right) $ as given by
eqs(\ref{so}).\newline
Notice that invariance of the effective potential $\mathcal{V}_{eff}$ under
the electric/magnetic duality symmetry between the charges of the black
branes and their duals is captured by the relation $(M,\sigma )\rightarrow
\left( E,-\sigma \right) $ with M standing form the magnetic charges and E
for electric ones.

\subsection{Attractor eqs}

These are given as usual by minimizing the effective potential with respect
to the field moduli $\sigma $ and $L_{\Lambda \Upsilon }$;
\begin{equation}
\begin{tabular}{lllll}
$\frac{\partial \mathcal{V}_{eff}}{\partial \sigma }$ & $=0$ & , & $\frac{%
\partial \mathcal{V}_{eff}}{\partial L_{\Lambda \Upsilon }}$ & $=0$%
\end{tabular}
\label{at}
\end{equation}%
by taking into account the constraint relation $L^{T}\eta L=\eta $. This
constraint relation may be implemented in the effective potential by using
the Lagrange multiplier method; for technical details see \textrm{\cite{6D} }%
developed for the case of black attractors in \emph{6D} supergravity. We
also need to compute the Hessian matrix
\begin{equation}
\begin{tabular}{lllll}
$\frac{\partial ^{2}\mathcal{V}_{eff}}{\partial \sigma ^{2}}=0$ & , & $\frac{%
\partial ^{2}\mathcal{V}_{eff}}{\partial \sigma \partial L_{\Lambda \Upsilon
}}=0$ & , & $\frac{\partial ^{2}\mathcal{V}_{eff}}{\partial L_{\Lambda
\Upsilon }\partial L_{\Gamma \Sigma }}=0$%
\end{tabular}%
\end{equation}%
which needs to be positive definite for stable solutions.

\emph{Computing} $\partial \mathcal{V}_{eff}/\partial \sigma =0$\newline
Now, using the fact that $X^{i},$ $\tilde{X}_{i},$ $Y^{a}$, $\tilde{Y}_{a}$
are eigenvectors of $\frac{\partial }{\partial \sigma }$ ; i.e $\frac{%
\partial X^{i}}{\partial \sigma }=X^{i}$, $\frac{\partial Y^{a}}{\partial
\sigma }=Y^{a}$, $\frac{\partial \tilde{X}_{i}}{\partial \sigma }=-\tilde{X}%
_{i}$, $\frac{\partial \tilde{Y}_{a}}{\partial \sigma }=-\tilde{Y}_{a}$, the
extremization with respect to the dilaton $\sigma $ gives,%
\begin{equation}
\begin{tabular}{lll}
$0=$ & $+\dsum\limits_{i,j=1}^{2}\left( X^{i}\delta _{ij}X^{j}-\tilde{X}%
_{i}\delta ^{ij}\tilde{X}_{j}\right) $ &  \\
& $+\dsum\limits_{a,b=1}^{N}\left( Y^{a}\delta _{ab}Y^{b}-\tilde{Y}%
_{a}\delta ^{ab}\tilde{Y}_{b}\right) $ &  \\
& $+2\left( e^{4\sigma }p_{0}^{2}-e^{-4\sigma }q_{0}^{2}\right) $ & .%
\end{tabular}
\label{sig}
\end{equation}%
There are different ways to solve this attractor eq; one of them is to cast
it as follows%
\begin{equation}
\begin{tabular}{lll}
$X^{i}\delta _{ij}X^{j}-\tilde{X}_{i}\delta ^{ij}\tilde{X}_{j}$ & $=0$ & ,
\\
$Y^{a}\delta _{ab}Y^{b}-\tilde{Y}_{a}\delta ^{ab}\tilde{Y}_{b}$ & $=0$ & ,
\\
$e^{4\sigma }p_{0}^{2}-e^{-4\sigma }q_{0}^{2}$ & $=0$ & ,%
\end{tabular}
\label{X}
\end{equation}%
where summation on repeated indices is understood. An other way is to
compensate the terms with $X^{i},$ $\tilde{X}_{i}$ with the terms with $%
Y^{b},$ $\tilde{Y}_{a}$ as follows,%
\begin{equation}
\begin{tabular}{lll}
$X^{i}\delta _{ij}X^{j}+Y^{a}\delta _{ab}Y^{b}$ & $=0$ & , \\
$\tilde{X}_{i}\delta ^{ij}\tilde{X}_{j}+\tilde{Y}_{a}\delta ^{ab}\tilde{Y}%
_{b}$ & $=0$ & , \\
$e^{4\sigma }p_{0}^{2}-e^{-4\sigma }q_{0}^{2}$ & $=0$ & ,%
\end{tabular}
\label{YY}
\end{equation}%
or like%
\begin{equation}
\begin{tabular}{lll}
$X^{i}\delta _{ij}X^{j}-\tilde{Y}_{a}\delta ^{ab}\tilde{Y}_{b}$ & $=0$ & ,
\\
$\tilde{X}_{i}\delta ^{ij}\tilde{X}_{j}-Y^{a}\delta _{ab}Y^{b}$ & $=0$ & ,
\\
$e^{4\sigma }p_{0}^{2}-e^{-4\sigma }q_{0}^{2}$ & $=0$ & .%
\end{tabular}
\label{ZZ}
\end{equation}%
Further solutions are obtained by compensating $X^{i},$ $\tilde{X}_{i}$, $%
Y^{b},$ $\tilde{Y}_{a}$ with $e^{4\sigma }p_{0}^{2}$ and $e^{-4\sigma
}q_{0}^{2}$; for instance as follows:
\begin{equation}
\begin{tabular}{llll}
$X^{i}\delta _{ij}X^{j}$ & $=$ & $e^{-4\sigma }q_{0}^{2}$ & , \\
$e^{4\sigma }p_{0}^{2}$ & $=$ & $\tilde{X}_{i}\delta ^{ij}\tilde{X}_{j}$ & ,
\\
$Y^{a}\delta _{ab}Y^{b}$ & $=$ & $\tilde{Y}_{a}\delta ^{ab}\tilde{Y}_{b}$ & .%
\end{tabular}
\label{YZ}
\end{equation}%
We will give some explicit examples later on.\newline
Substituting the dressed central charges by their field expressions back
into (\ref{X}), we get the following attractor eqs%
\begin{equation}
\begin{tabular}{llll}
$e^{2\sigma }P^{\Lambda }L_{\Lambda }^{i}\delta _{ij}L_{\Upsilon
}^{j}P^{\Upsilon }-e^{-2\sigma }Q_{\Lambda }\left( L^{-1}\right)
_{i}^{\Lambda }\delta ^{ij}\left( L^{-1}\right) _{j}^{\Upsilon }Q_{\Upsilon
} $ & $=$ & $0$ & , \\
$e^{2\sigma }P^{\Lambda }L_{\Lambda }^{a}\delta _{ab}L_{\Upsilon
}^{b}P^{\Upsilon }-e^{-2\sigma }Q_{\Lambda }\left( L^{-1}\right)
_{a}^{\Lambda }\delta ^{ab}\left( L^{-1}\right) _{b}^{\Upsilon }Q_{\Upsilon
} $ & $=$ & $0$ & , \\
$\left( e^{4\sigma }p_{0}^{2}-e^{-4\sigma }q_{0}^{2}\right) $ & $=$ & $0$ & .%
\end{tabular}%
\end{equation}%
Similar attractor eqs may be written down for the other cases given above.

\emph{Computing }$\delta _{L}\mathcal{V}_{eff}=0$\newline
The extremization of the effective potential of the black attractors with
respect to the field matrix $L_{\Lambda \Upsilon }$ is some how lengthy.
Below, we give the main steps by using the expression of $\mathcal{V}_{eff}$
in terms of $X^{i},$ $\tilde{X}_{i},$ $Y^{a}$, $\tilde{Y}_{a}$. First, we
have%
\begin{equation}
\begin{tabular}{ll}
$\delta _{L_{\Lambda \Upsilon }}\mathcal{V}_{eff}=$ & $+2\dsum%
\limits_{i,j=1}^{2}\left[ \left( \delta _{L_{\Lambda \Upsilon }}X^{i}\right)
\delta _{ij}X^{j}+\left( \delta _{L_{\Lambda \Upsilon }}\tilde{X}_{i}\right)
\delta ^{ij}\tilde{X}_{j}\right] $ \\
& $+2\dsum\limits_{a,b=1}^{N}\left[ \left( \delta _{L_{\Lambda \Upsilon
}}Y^{a}\right) \delta _{ab}Y^{b}+\left( \delta _{L_{\Lambda \Upsilon }}%
\tilde{Y}_{a}\right) \delta ^{ab}\tilde{Y}_{b}\right] $%
\end{tabular}
\label{ef}
\end{equation}%
where $\delta _{L_{\Lambda \Upsilon }}X^{i}$ and so on are the variation of
the dressed central charges with respect to the field matrix $L_{\Lambda
\Upsilon }$. These variations may be nicely expressed in terms of the Maurer
-Cartan 1-form
\begin{equation}
\begin{tabular}{ll}
$\Omega =dLL^{-1}=-L\left( dL^{-1}\right) $ & ,%
\end{tabular}%
\end{equation}%
of the orthogonal group $SO\left( 2,N\right) $. Indeed, denoting $%
X^{i}=e^{\sigma }L_{\Lambda }^{i}P^{\Lambda }$ in a condensed manner as $%
X^{i}=e^{\sigma }\left( L.P\right) ^{i}$ and similarly for the other dressed
central charges, the variation with respect to $L$ reads as $\delta
X^{i}=e^{\sigma }\left( \delta L.P\right) ^{i}$. Now inserting the relation $%
L^{-1}L=I$, we get $\delta X=e^{\sigma }\left( \delta L.L^{-1}.LP\right)
^{i} $ where we recognize the $\Omega $ term. Doing the same for the other
dressed central charges, we end with the following result:
\begin{equation}
\begin{tabular}{llll}
$\delta X^{i}=\Omega _{k}^{i}.X^{k}+\Omega _{c}^{i}.Y^{c}$ & , & $\delta
Y^{a}=\Omega _{k}^{a}.X^{k}+\Omega _{c}^{a}.Y^{c}$ &  \\
$\delta \tilde{X}_{i}=-\Omega _{i}^{k}.\tilde{X}_{k}-\Omega _{i}^{c}.\tilde{Y%
}_{c}$ & , & $\delta \tilde{Y}_{a}=-\Omega _{a}^{k}.\tilde{X}_{k}-\Omega
_{a}^{c}.\tilde{Y}_{c}$ &
\end{tabular}
\label{tran}
\end{equation}%
Putting back into (\ref{ef}), we get the vanishing condition of $\delta _{L}%
\mathcal{V}_{eff}$
\begin{equation}
\begin{tabular}{llll}
$\left( X\Omega X-\tilde{X}\Omega \tilde{X}\right) $ & $+$ & $\left( X\Omega
Y-\tilde{X}\Omega \tilde{Y}\right) $ &  \\
$+\left( Y\Omega Y-\tilde{Y}\Omega \tilde{Y}\right) $ & $+$ & $\left(
Y\Omega X-\tilde{Y}\Omega \tilde{X}\right) $ & $=0$%
\end{tabular}
\label{L}
\end{equation}%
from which we can learn the associated attractor eqs. In this relation, the
condensed terms are as follows%
\begin{equation}
\begin{tabular}{llllll}
$X\Omega X$ & $=+X^{j}\Omega _{jk}X^{k}$ & , & $\tilde{X}\Omega \tilde{X}$ &
$=+\tilde{X}_{j}\Omega ^{jk}\tilde{X}_{k}$ &  \\
$Y\Omega Y$ & $=-Y^{b}\Omega _{bc}Y^{c}$ & , & $\tilde{Y}\Omega \tilde{Y}$ &
$=-\tilde{Y}_{b}\Omega ^{bc}\tilde{Y}_{c}$ &  \\
$X\Omega Y$ & $=+X^{j}\Omega _{jc}Y^{c}$ & , & $\tilde{X}\Omega \tilde{Y}$ &
$=+\tilde{X}_{j}\Omega ^{jc}\tilde{Y}_{c}$ &  \\
$Y\Omega X$ & $=-Y^{b}\Omega _{bk}X^{k}$ & , & $\tilde{Y}\Omega \tilde{X}$ &
$=-\tilde{Y}_{b}\Omega ^{bk}\tilde{X}_{k}$ &
\end{tabular}%
\end{equation}%
where the i,j indices are raised and lowered by $\delta _{ij}$ and $\delta
^{ij}$ while the indices a, b are raised and lowered by $-\delta _{ab}$ and $%
-\delta ^{ab}$. Notice also that we have
\begin{equation}
\begin{tabular}{lll}
$\Omega _{\Lambda \Upsilon }^{\left( 2+N,2+N\right) }=$ & $\left(
\begin{array}{cc}
\Omega _{ij}^{\left( 2,2\right) } & \Omega _{ib}^{\left( 2,N\right) } \\
\Omega _{aj}^{\left( N,2\right) } & \Omega _{ab}^{\left( N,N\right) }%
\end{array}%
\right) $ &
\end{tabular}%
\end{equation}%
Notice as well that the attractor eqs of the black attractors (\ref{at}) are
given by eqs(\ref{sig},\ref{L}); a class of solutions of these eqs are given
below.

\subsection{Solutions of attractor eqs}

We first solve the attractor eq $\partial \mathcal{V}_{eff}/\partial \sigma
=0$ (\ref{sig}) allowing to fix the dilaton in terms of the electric and
magnetic bare charges. Then, we consider the case of the attractor eqs $%
\partial \mathcal{V}_{eff}/\partial L_{\Lambda \Upsilon }=0$.

\subsubsection{Solving $\partial \mathcal{V}_{eff}/\partial \protect\sigma %
=0 $}

Eq(\ref{sig}) may be solved in several ways:\newline
(\textbf{1}) \emph{no D-brane fluxes:} $P^{\Lambda }=0,$ $Q_{\Lambda }=0$
\newline
This configuration corresponds to $X^{i}=0,Y^{a}=0$; $\tilde{X}_{i}=0,$ $%
\tilde{Y}_{a}=0$. Substituting, eq(\ref{sig}) reduces to $\left( e^{4\sigma
}p_{0}^{2}-e^{-4\sigma }q_{0}^{2}\right) =0$ whose solution is
\begin{equation}
\begin{tabular}{ll}
$\sigma _{0}=\frac{-1}{4}\ln \frac{p_{0}}{q_{0}}$ &
\end{tabular}
\label{s}
\end{equation}%
giving the value of the dilaton in terms of the magnetic charge of the
string and the electric charge of the 3-brane. The near horizon geometry of
this black attractor is given by $AdS_{3}\times S^{5}$ and $AdS_{5}\times
S^{3}$ depending on the values of the magnetic and electric charges. Notice
that for $p_{0}=0,$ $\sigma _{0}\rightarrow +\infty $ while for $q_{0}=0,$ $%
\sigma _{0}\rightarrow -\infty .$\newline
(\textbf{2}) \emph{general solutions}\newline
These solutions correspond to compensate the contributions coming from the
electric and magnetic sectors as in eqs(\ref{X},\ref{YY},\ref{ZZ},\ref{YZ}).
As an example, we consider the case%
\begin{equation}
\begin{tabular}{llll}
$X^{i}\delta _{ij}X^{j}$ & $=$ & $\tilde{X}_{i}\delta ^{ij}\tilde{X}_{j}$ & ,
\\
$Y^{a}\delta _{ab}Y^{b}$ & $=$ & $\tilde{Y}_{a}\delta ^{ab}\tilde{Y}_{b}$ & ,
\\
$\sigma _{0}$ & $=$ & $\frac{1}{4}\ln \frac{q_{0}}{p_{0}}$ & .%
\end{tabular}%
\end{equation}%
which may be solved in four ways by taking the dressed central charges as
follows:%
\begin{equation}
\begin{tabular}{llllllllll}
$\left( i\right) $ & $:$ & $X^{i}=+z^{i}$ & $,$ & $\tilde{X}%
_{j}=+z^{i}\delta _{ij}$ & $,$ & $Y^{a}=+y^{a}$ & $,$ & $\tilde{Y}%
_{b}=+y^{a}\delta _{ab}$ &  \\
$\left( ii\right) $ & $:$ & $X^{i}=+z^{i}$ & $,$ & $\tilde{X}%
_{j}=+z^{i}\delta _{ij}$ & $,$ & $Y^{a}=+y^{a}$ & $,$ & $\tilde{Y}%
_{b}=-y^{a}\delta _{ab}$ &  \\
$\left( iii\right) $ & $:$ & $X^{i}=-z^{i}$ & $,$ & $\tilde{X}%
_{j}=-z^{i}\delta _{ij}$ & $,$ & $Y^{a}=-y^{a}$ & $,$ & $\tilde{Y}%
_{b}=+y^{a}\delta _{ab}$ &  \\
$\left( iv\right) $ & $:$ & $X^{i}=-z^{i}$ & $,$ & $\tilde{X}%
_{j}=-z^{i}\delta _{ij}$ & $,$ & $Y^{a}=-y^{a}$ & $,$ & $\tilde{Y}%
_{b}=-y^{a}\delta _{ab}$ &
\end{tabular}
\label{xy}
\end{equation}%
and $\sigma _{0}$ as before and $z^{i},$ $y^{a}$ some constants. The above
relations may also written as follows%
\begin{equation}
\begin{tabular}{lll}
$\left( L^{-1}\right) _{i}^{\Lambda }Q_{\Lambda }$ & $=\pm e^{-2\sigma
_{0}}\left( L\right) _{\Upsilon }^{j}P^{\Upsilon }\delta _{ij}$ & , \\
$\left( L^{-1}\right) _{b}^{\Lambda }Q_{\Lambda }$ & $=\pm e^{-2\sigma
_{0}}\left( L\right) _{\Upsilon }^{a}P^{\Upsilon }\delta _{ab}$ & , \\
$L^{T}\eta ^{SO\left( 2,N\right) }L$ & $=\eta ^{SO\left( 2,N\right) }$ & ,%
\end{tabular}%
\end{equation}%
with $e^{-2\sigma _{0}}$ given by eq(\ref{s}) and whose solutions allow to
express the field matrix $L_{\Lambda \Upsilon }$ in terms of $Q_{\Lambda },$
$P^{\Upsilon }$ as well as $\frac{p_{0}}{q_{0}}$.\newline
Notice that the moduli space of solutions of (\ref{xy}) depends on the
arbitrary values $z^{i}$ and $y^{a}$. For instance taking
\begin{equation}
\begin{tabular}{llll}
$\tilde{Y}_{b}=\left(
\begin{array}{c}
y_{1} \\
0 \\
\vdots \\
0%
\end{array}%
\right) $ & , & $Y^{a}\delta _{ab}=\left(
\begin{array}{c}
\pm y_{1} \\
0 \\
\vdots \\
0%
\end{array}%
\right) $ &
\end{tabular}%
\end{equation}%
the $SO\left( 2\right) \times SO\left( N\right) $ symmetry of the moduli
space get reduced down to $SO\left( N-1\right) $. Generic solutions read as
follows%
\begin{equation}
\begin{tabular}{llll}
$\tilde{Y}_{b}=\left(
\begin{array}{c}
y_{1} \\
\vdots \\
y_{n} \\
0 \\
\vdots \\
0%
\end{array}%
\right) $ & , & $Y^{a}\delta _{ab}=\left(
\begin{array}{c}
\pm y_{1} \\
\vdots \\
\pm y_{n} \\
0 \\
\vdots \\
0%
\end{array}%
\right) $ &
\end{tabular}%
\end{equation}%
and have a $SO\left( N-n\right) $ symmetry.

\subsubsection{Solving $\partial \mathcal{V}_{eff}/\partial \protect\sigma %
=0 $ and $\partial \mathcal{V}_{eff}/\partial L_{\Lambda \Upsilon }=0$}

We give here below two classes of solutions; others solutions classified by
the $SO\left( N-n\right) $ symmetries can be also written down.

\emph{Class I}\newline
The first class of solutions of the attractor eqs(\ref{at},\ref{sig},\ref{L}%
) is obtained by putting eqs(\ref{s}) and (\ref{xy}) back into eq(\ref{L});
then cast it as follows:%
\begin{equation}
\begin{tabular}{llllll}
$X\Omega X$ & $-$ & $\tilde{X}\Omega \tilde{X}$ & $=$ & $0$ &  \\
$Y\Omega Y$ & $-$ & $\tilde{Y}\Omega \tilde{Y}$ & $=$ & $0$ &  \\
$X\Omega Y$ & $-$ & $\tilde{X}\Omega \tilde{Y}$ & $=$ & $0$ &  \\
$Y\Omega X$ & $-$ & $\tilde{Y}\Omega \tilde{X}$ & $=$ & $0$ &
\end{tabular}
\label{zz}
\end{equation}%
Taking into account eqs(\ref{s}, \ref{xy}) solving $\partial \mathcal{V}%
_{eff}/\partial \sigma =0$, it is not difficult to see that the solutions of
eq(\ref{zz}) are classified as given below%
\begin{equation}
\begin{tabular}{llllll}
$\left( i\right) $ & : & $X=+\tilde{X}=\left(
\begin{array}{c}
z_{1} \\
z_{2}%
\end{array}%
\right) $ & $,$ & $Y=+\tilde{Y}=\left(
\begin{array}{c}
w_{1} \\
\vdots \\
w_{N}%
\end{array}%
\right) $ &  \\
&  &  &  &  &  \\
$\left( ii\right) $ & : & $X=-\tilde{X}=\left(
\begin{array}{c}
z_{1}^{\prime } \\
z_{2}^{\prime }%
\end{array}%
\right) $ & $,$ & $Y=-\tilde{Y}=\left(
\begin{array}{c}
w_{1}^{\prime } \\
\vdots \\
w_{N}^{\prime }%
\end{array}%
\right) $ &
\end{tabular}%
\end{equation}%
together with $\sigma _{0}=\frac{1}{4}\ln \frac{q_{0}}{p_{0}}$ and where $%
z_{i},$ $w_{a}$ and $z_{i}^{\prime },$ $w_{a}^{\prime }$ are some constant
numbers. \newline
Notice that the terms $X\Omega X=X^{i}\Omega _{ij}X^{j}$ and $Y\Omega
Y=Y^{a}\Omega _{ab}Y^{b}$ are symmetric quadratic forms; so there no
contribution coming from the antisymmetric parts $\Omega _{\left[ ij\right]
}=\left( \Omega _{ij}-\Omega \right) _{ji}$ and $\Omega _{\left[ ab\right]
}=\Omega _{ab}-\Omega _{ba}$\ of the Cartan-Maurer forms. This property
captures precisely the $SO\left( 2\right) \times SO\left( N\right) $
symmetry of the moduli space (\ref{het}-\ref{pm}) of non chiral \emph{8D }$%
\mathcal{N}=1$ supergravity.\newline
Notice also that for arbitrary values of $z_{i},$ $w_{a}$ and $z_{i}^{\prime
},$ $w_{a}^{\prime }$ the symmetry group $SO\left( 2\right) \times SO\left(
N\right) $ of the effective potential is completely broken. \ The other
possibilities where some of the parameters are zero or identical, the $%
SO\left( 2\right) \times SO\left( N\right) $ symmetry of the moduli space is
broken down to a subgroup $G$ .

\emph{Class II}\newline
This class of solutions corresponds to solving the extremum of $\mathcal{V}%
_{eff}$ (\ref{L}) by compensating the $X^{i}$ and $\tilde{X}_{i}$ factors
with the $Y^{a}$ and $\tilde{Y}_{a}$ as in (\ref{YY}-\ref{ZZ}). A way to do
it is as follows:%
\begin{equation}
\begin{tabular}{llll}
$Y\Omega Y=\tilde{X}\Omega \tilde{X}$ & $,$ & $\tilde{Y}\Omega \tilde{Y}%
=X\Omega X$ &  \\
$Y\Omega X=\tilde{X}\Omega \tilde{Y}$ & $,$ & $\tilde{Y}\Omega \tilde{X}%
=X\Omega Y$ &
\end{tabular}%
\end{equation}%
In this solution, the term $Y\Omega Y$ (resp $\tilde{Y}\Omega \tilde{Y}$) is
compensated by $\tilde{X}\Omega \tilde{X}$ (resp $X\Omega X$); this
corresponds to first breaking the moduli space $SO\left( 2\right) \times
SO\left( N\right) $ subsymmetry as%
\begin{equation}
\begin{tabular}{llll}
$SO\left( 2\right) \times SO\left( N\right) $ & $\rightarrow $ & $SO\left(
2\right) \times SO\left( 2\right) \times SO\left( N-2\right) $ & ,%
\end{tabular}%
\end{equation}%
then compensate the terms associated with the two $SO\left( 2\right) $
factor. An explicit solution is given by:%
\begin{equation}
\begin{tabular}{ll}
$X=\left(
\begin{array}{c}
z_{1} \\
z_{2}%
\end{array}%
\right) ,$ & $\tilde{X}=\left(
\begin{array}{c}
\tilde{z}_{1} \\
\tilde{z}_{2}%
\end{array}%
\right) ,$ \\
&  \\
$Y=\left(
\begin{array}{c}
\tilde{z}_{1} \\
\tilde{z}_{2} \\
0 \\
\vdots \\
0%
\end{array}%
\right) ,$ & $\tilde{Y}=\left(
\begin{array}{c}
z_{1} \\
z_{2} \\
0 \\
\vdots \\
0%
\end{array}%
\right) $%
\end{tabular}
\label{xxx}
\end{equation}%
Other configurations with symmetries $SO\left( N-m\right) $ with $m=3,...N$
may be also written down; one of them is given by $X^{i}$ and $\tilde{X}_{i}$
as in (\ref{xxx}) and $Y^{a}$ and $\tilde{Y}_{a}$ like,%
\begin{equation}
\begin{tabular}{ll}
$Y=\left(
\begin{array}{c}
\tilde{z}_{1} \\
\tilde{z}_{2} \\
w \\
0 \\
\vdots \\
0%
\end{array}%
\right) ,$ & $\tilde{Y}=\left(
\begin{array}{c}
z_{1} \\
z_{2} \\
w \\
0 \\
\vdots \\
0%
\end{array}%
\right) $%
\end{tabular}
\label{yyy}
\end{equation}%
with $SO\left( N-3\right) $ symmetry.

\section{Intersecting attractors}

Following \textrm{\cite{70,7A}}, one should distinguish two main classes of
black p-brane solutions in higher dimensional supergravity. In non chiral
\emph{8D} $\mathcal{N}=1$ supergravity we are considering here, these are:%
\newline
(\textbf{1}) the standard black p- brane solutions based on $AdS_{2+p}\times
S^{6-p}$ with $p=0,1,3,4$, whose basic features have been given above.%
\newline
(\textbf{2}) the intersecting attractors with the typical near horizon
geometries
\begin{equation}
AdS_{2+p}\times S^{m}\times M^{6-p-m}  \label{aa}
\end{equation}%
where $S^{m}$ is the real m-sphere and $M^{n}$ stands for some manifolds;
essentially a $n$-torus. Moreover, since there is no D2- brane flux in this
theory; these geometries are restricted to black hole and black string
geometries as well as their duals. As such, we have:\newline
(\textbf{a}) $AdS_{3}\times S^{m}\times M^{5-m}$\newline
(\textbf{b}) $AdS_{2}\times S^{m}\times M^{6-m}$ \newline
The novelty with these geometries is that they allow the two following
features: \newline
(\textbf{i}) a variety of irreducible sub-manifolds that support various
kinds of branes and so a rich spectrum of electric and magnetic charges;
\newline
(\textbf{ii}) non trivial intersections between $p_{i}$-/$p_{j}$- cycles of (%
\ref{aa}) leading to intersecting (BPS and non BPS) attractors. \newline
To illustrate the first point, consider the example of the two compact
manifolds $S^{m+n}$ and $\mathcal{M}^{m+n}=S^{m}\times T^{n}$. While the
sphere $S^{m+n}$ supports only charges of $\left( m+n-2\right) $-brane
charges%
\begin{equation}
\begin{tabular}{llll}
$\mathcal{F}_{n+m}=g$ $\varpi _{n+m}$ & , & $g=\int_{S^{m+n}}\mathcal{F}%
_{n+m}$ & ,%
\end{tabular}%
\end{equation}%
and no $\left( m-1\right) $- brane nor others, the manifold $S^{m}\times
T^{n}$ allows however many possibilities. It has several irreducible $k_{i}$%
- cycles that support, in addition to $\left( m+n-2\right) $-branes, other
kinds; in particular $n$ types of $\left( m-1\right) $-branes with charges
given by,%
\begin{equation}
\begin{tabular}{llll}
$g^{a}=\int_{\mathcal{C}_{m+1}^{(a)}}\mathcal{F}_{m+1}$ & $,$ & $\mathcal{F}%
_{m+1}=\sum\limits_{a}g^{a}\varpi _{_{m+1|a}}$ & , \\
$\int_{\mathcal{C}_{m+1}^{(a)}}\varpi _{_{m+1|b}}=\delta _{b}^{a}$ & , & $%
a=1,...,n$ & ,%
\end{tabular}%
\end{equation}%
with
\begin{equation}
\begin{tabular}{llll}
$\mathcal{C}_{m+1}^{(a)}=\dbigcup\limits_{a=1}^{n}\left( S_{{\scriptsize a}%
}^{1}\times S^{m}\right) $ & , & $T^{n}=\dbigotimes\limits_{a=1}^{n}S_{%
{\scriptsize a}}^{1}$ & .%
\end{tabular}%
\end{equation}%
The branes may be imagined as filling the fiber $\mathsf{F}_{m-1}^{(a)}$ of
these cycles $\mathcal{C}_{m+1}^{(a)}$ thought of in terms of the fibration $%
\mathcal{C}_{m+1}^{(a)}\sim \mathsf{F}_{m-1}^{(a)}\times S^{2}$ with field
strength
\begin{equation}
\begin{tabular}{lll}
$\mathcal{F}_{m+1}$ & $=\mathbf{\beta }_{S^{2}}\wedge \left(
\sum\limits_{a}g^{a}\mathbf{\beta }_{\mathsf{F}_{m-1}^{(a)}}\right) $ &
\end{tabular}%
\end{equation}%
Using the anzats of \textrm{\cite{8A}}, we focus below on the study of
various examples of these typical horizon geometries and work out new and
explicit solutions regarding intersecting attractors in the case of non
chiral \emph{8D} $\mathcal{N}=1$ supergravity. As the solutions are very
technical, we will concentrate on drawing their main lines and giving the
results.

\subsection{Geometries with $AdS_{3}$ and $AdS_{5}$ factors}

We distinguish several $AdS_{3}\times S^{m}\times M^{5-m}$ and their $%
AdS_{5}\times S^{m}\times M^{3-m}$ duals geometries; in particular: \newline
(a) $AdS_{3}\times S^{3}\times T^{2}$ with volume forms $\mathbf{\alpha }%
_{AdS_{3}},$ $\mathbf{\beta }_{S^{3}}$ and $\mathbf{\beta }_{T^{2}}$,\newline
(b) $AdS_{3}\times S^{2}\times T^{3}$ with volume forms $\mathbf{\alpha }%
_{AdS_{3}},$ $\mathbf{\beta }_{S^{2}}$ and $\mathbf{\beta }_{T^{3}}$,\newline
(c) $AdS_{3}\times S^{4}\times S^{1}$ with volume forms $\mathbf{\alpha }%
_{AdS_{3}},$ $\mathbf{\beta }_{S^{4}}$ and $\mathbf{\beta }_{S^{1}}.$\newline
Below, we study the two first ones.

\subsubsection{$AdS_{3}\times S^{3}\times T^{2}$}

On the geometry $AdS_{3}\times S^{3}\times T^{2}$ there is no irreducible
2-cycle nor irreducible 6-cycle that support the fluxes emanating from the
D0 and D6- branes. As such the black attractor is given by,%
\begin{equation}
\begin{tabular}{l|l|l|}
\cline{2-3}
& p-branes & $\left( 4-p\right) $- branes \\ \hline
\multicolumn{1}{|l|}{$p=0$} & $\mathcal{F}_{2}^{\Lambda }=0$ & $\mathcal{%
\tilde{F}}_{6|\Lambda }=0$ \\ \hline
\multicolumn{1}{|l|}{$p=1$} & $\mathcal{F}_{3}=p^{0}\mathbf{\beta }_{S^{3}}$
& $\mathcal{\tilde{F}}_{5}=q_{0}$ $\mathbf{\alpha }_{AdS_{3}}\wedge $ $\beta
_{T^{2}}$ \\ \hline
\end{tabular}%
\end{equation}%
from which we read the following effective potential,%
\begin{equation}
\begin{tabular}{ll}
$\mathcal{V}_{eff}=$ & $e^{4\sigma }p_{0}^{2}+e^{-4\sigma }q_{0}^{2}$%
\end{tabular}%
\end{equation}%
The extremization of this potential with respect to the dilaton leads to%
\begin{equation}
e^{4\sigma }p_{0}^{2}-e^{-4\sigma }q_{0}^{2}=0  \label{dil}
\end{equation}%
The solving of the above equation is given by $\sigma _{0}=\frac{1}{4}\ln
\frac{q_{0}}{p_{0}}$; it fixes the dilaton $\sigma _{0}$ at horizon in terms
of the magnetic charge of the black string and the electric charge of the
black 3-brane.

\subsubsection{$AdS_{3}\times S^{2}\times T^{3}$}

In this geometry which involve the volume forms $\mathbf{\alpha }_{\text{%
{\small AdS}}_{3}}$, $\mathbf{\beta }_{S^{2}}$, $\mathbf{\beta }_{T^{3}}$,
the non vanishing field strength charges are given by%
\begin{equation}
\begin{tabular}{l|l|l|}
\cline{2-3}
& p-branes & $\left( 4-p\right) $- branes \\ \hline
\multicolumn{1}{|l|}{$p=0$} & $\mathcal{F}_{2}^{\Lambda }=P^{\Lambda }$ $%
\mathbf{\beta }_{S^{2}}$ & $\mathcal{\tilde{F}}_{6|\Upsilon }=Q_{\Upsilon }$
$\left( \mathbf{\alpha }_{\text{{\small AdS}}_{3}}{\small \wedge }\mathbf{%
\beta }_{T^{3}}\right) $ \\ \hline
\multicolumn{1}{|l|}{$p=1$} & $\mathcal{F}_{3}=p^{0}$ $\mathbf{\alpha }_{%
\text{{\small AdS}}_{3}}$ & $\mathcal{\tilde{F}}_{5}=q_{0}$ $\left( \mathbf{%
\beta }_{S^{2}}{\small \wedge }\mathbf{\beta }_{T^{3}}\right) $ \\ \hline
\end{tabular}%
\end{equation}%
where $\Lambda =\left( i,a\right) $ with $i=1,2$ and $a=1,...,N.$\newline
The effective potential $\mathcal{V}_{eff}$ of these black attractor
configuration reads as follows,%
\begin{equation}
\begin{tabular}{ll}
$\mathcal{V}_{eff}=$ & $+\dsum\limits_{i,j=1}^{2}\left( X^{i}\delta
_{ij}X^{j}+\tilde{X}_{i}\delta ^{ij}\tilde{X}_{j}\right) +\left( e^{4\sigma
}p_{0}^{2}+e^{-4\sigma }q_{0}^{2}\right) $ \\
& $+\dsum\limits_{a,b=1}^{N}\left( Y^{a}\delta _{ab}Y^{b}+\tilde{Y}%
_{a}\delta ^{ab}\tilde{Y}_{b}\right) $%
\end{tabular}
\label{vv}
\end{equation}%
where the first term, which we write as $XX+\tilde{X}\tilde{X},$ is
invariant under SO$\left( 2\right) $ and the term $YY+\tilde{Y}\tilde{Y}$ is
invariant under $SO\left( N\right) $. The extremization of $\mathcal{V}%
_{eff} $ gives,%
\begin{equation}
\begin{tabular}{llll}
$\left( X\Omega X-\tilde{X}\Omega \tilde{X}\right) +\left( X\Omega Y-\tilde{X%
}\Omega \tilde{Y}\right) $ &  &  &  \\
$\left( Y\Omega Y-\tilde{Y}\Omega \tilde{Y}\right) +\left( Y\Omega X-\tilde{Y%
}\Omega \tilde{X}\right) $ & $=$ & $0$ &  \\
$e^{4\sigma }p_{0}^{2}-e^{-4\sigma }q_{0}^{2}$ & $=$ & $0$ &
\end{tabular}%
\end{equation}%
where $\Omega $ is the Maurer Cartan 1-form of $SO\left( 2,N\right) $
introduced previously. \newline
The solutions of these attractor eqs may be realized in various ways; one of
them is given by the following:%
\begin{equation}
\begin{tabular}{lllll}
$X=\pm \tilde{X}$ & , & $Y=\pm \tilde{Y}$ & , & $\sigma _{0}=\frac{-1}{4}\ln
\frac{p_{0}}{q_{0}}$%
\end{tabular}%
\end{equation}%
These solutions correspond to diverse intersecting configurations composed
of a black hole, a black 4-brane, a black string,and a black 3-brane.\newline
Moreover, using eqs(\ref{tran}), we compute the following the Hessian matrix%
\begin{equation}
\begin{tabular}{ll}
$\delta \delta \mathcal{V}_{eff}=$ & $+16\left( e^{4\sigma
}p_{0}^{2}+e^{-4\sigma }q_{0}^{2}\right) $ \\
& $+4\left( X\Omega ^{2}X-\tilde{X}\Omega ^{2}\tilde{X}+X\Omega ^{2}Y-\tilde{%
X}\Omega ^{2}\tilde{Y}\right) $ \\
& $+4\left( Y\Omega ^{2}Y-\tilde{Y}\Omega ^{2}\tilde{Y}+Y\Omega ^{2}X-\tilde{%
Y}\Omega ^{2}\tilde{X}\right) $ \\
& $+\left[ X\left( \delta \Omega \right) Y-\tilde{X}\left( \delta \Omega
\right) \tilde{Y}\right] \ +\left[ Y\left( \delta \Omega \right) Y-\tilde{Y}%
\left( \delta \Omega \right) \tilde{Y}\right] $ \\
& $+\left[ Y\left( \delta \Omega \right) X-\tilde{Y}\left( \delta \Omega
\right) \tilde{X}\right] +\left[ X\left( \delta \Omega \right) X-\tilde{X}%
\left( \delta \Omega \right) \tilde{X}\right] $%
\end{tabular}%
\end{equation}%
For the case $X=+\tilde{X}$, $Y=+\tilde{Y}$ and $X=-\tilde{X}$, $Y=-\tilde{Y}
$ the Hessian reduces to%
\begin{equation}
\begin{tabular}{ll}
$\delta \delta \mathcal{V}_{eff}=$ & $+16\left( e^{4\sigma
}p_{0}^{2}+e^{-4\sigma }q_{0}^{2}\right) $%
\end{tabular}%
\end{equation}%
which a positive definite definite quantity; it vanishes for $p_{0}=q_{0}=0$%
; that is no black string nor 3-brane. This is clearly seen by using the
identity $e^{4\sigma }p_{0}^{2}=e^{-4\sigma }q_{0}^{2}$ and replacing $%
e^{4\sigma }=\frac{q_{0}}{p_{0}}$, we get%
\begin{equation}
\begin{tabular}{llll}
$\delta \delta \mathcal{V}_{eff}=+32q_{0}p_{0}$ & , & $\mathcal{V}%
_{eff}=q_{0}p_{0}$ & .%
\end{tabular}%
\end{equation}

\subsection{Geometries with $AdS_{2}$ factor}

We study the following near horizon geometries. \newline
(\textbf{a}) $AdS_{2}\times S^{4}\times T^{2}$ with volume forms $\mathbf{%
\alpha }_{AdS_{2}},$ $\mathbf{\beta }_{S^{4}}$ and $\mathbf{\beta }_{T^{2}}$,%
\newline
(\textbf{b}) $AdS_{2}\times S^{3}\times T^{3}$ with volume forms $\mathbf{%
\alpha }_{AdS_{2}},$ $\mathbf{\beta }_{S^{3}}$ and $\mathbf{\beta }_{T^{3}}$,%
\newline
(\textbf{c}) $AdS_{2}\times S^{2}\times T^{4}$ with volume forms $\mathbf{%
\alpha }_{AdS_{2}},$ $\mathbf{\beta }_{S^{2}}$ and $\mathbf{\beta }_{T^{4}}.$

\subsubsection{$AdS_{2}\times S^{4}\times T^{2}$}

Using the various n-cycles of $AdS_{2}\times S^{4}\times T^{2}$ and the
corresponding n-forms that could live on, the general expressions of the
field strengths on this geometry reads as follows,%
\begin{equation}
\begin{tabular}{l|l|l|}
\cline{2-3}
& p-branes & $\left( \text{4-p}\right) $- branes \\ \hline
\multicolumn{1}{|l|}{$p=0$} & $\mathcal{F}_{2}^{\Lambda }=Q^{\Lambda }$ $%
\mathbf{\alpha }_{AdS_{2}}$ & $\mathcal{\tilde{F}}_{6|\Lambda }=P_{\Lambda }$
$\left( \mathbf{\beta }_{\emph{S}^{4}}\mathbf{\wedge \beta }_{\emph{T}%
^{2}}\right) $ \\ \hline
\multicolumn{1}{|l|}{$p=1$} & $\mathcal{F}_{3}=\sum%
\limits_{k=1}^{2}q_{k}^{0} $ $\left( \mathbf{\alpha }_{AdS_{2}}\mathbf{%
\wedge \alpha }_{S_{k}^{1}}\right) $ & $\mathcal{\tilde{F}}%
_{5}=\sum\limits_{k=1}^{2}p_{0k}\left( \mathbf{\beta }_{\emph{S}^{4}}\mathbf{%
\wedge \alpha }_{S_{k}^{1}}\right) $ \\ \hline
\end{tabular}%
\end{equation}%
where now the strings are charged electrically and the 3-branes
magnetically. \newline
The total effective potential $\mathcal{V}_{eff}$ associated with this
system is given as usual by the sum of the contribution of each extremal
black-brane. The attractor equations following from the extremization of $%
\mathcal{V}_{eff}$ are then given by:%
\begin{equation}
\begin{tabular}{llll}
$e^{4\sigma }\left( p_{01}^{2}+p_{02}^{2}\right) -e^{-4\sigma }\left(
q_{01}^{2}+q_{02}^{2}\right) $ & $=$ & $0$ &
\end{tabular}
\label{ss}
\end{equation}%
and
\begin{equation}
\begin{tabular}{llll}
$\left( X\Omega X-\tilde{X}\Omega \tilde{X}\right) +\left( X\Omega Y-\tilde{X%
}\Omega \tilde{Y}\right) $ &  &  &  \\
$\left( Y\Omega Y-\tilde{Y}\Omega \tilde{Y}\right) +\left( Y\Omega X-\tilde{Y%
}\Omega \tilde{X}\right) $ & $=$ & $0$ &
\end{tabular}
\label{st}
\end{equation}%
A class of solutions of (\ref{ss}-\ref{st}) is given by,
\begin{equation}
\begin{tabular}{llllll}
$\sigma _{0}=\frac{1}{8}\ln \left( \frac{q_{01}^{2}+q_{02}^{2}}{%
p_{01}^{2}+p_{02}^{2}}\right) $ & , & $X=\pm \tilde{X},$ & $Y=\pm \tilde{Y}$
&  & .%
\end{tabular}
\label{Y}
\end{equation}%
Other solutions like those given by eqs(\ref{xxx},\ref{yyy}) may be also
written down. Following the same method as before, we find in the case of (%
\ref{Y}) the following Hessian matrix at the horizon%
\begin{equation}
\delta \delta \mathcal{V}_{eff}=+32\sqrt{\left( q_{01}^{2}+q_{02}^{2}\right)
\left( p_{01}^{2}+p_{02}^{2}\right) }
\end{equation}

\subsubsection{$AdS_{2}\times S^{3}\times T^{3}$}

The general form of the field strengths on this geometry reads as,%
\begin{equation}
\begin{tabular}{l|l|l|}
\cline{2-3}
& p-branes & $\left( 4-p\right) $- branes \\ \hline
\multicolumn{1}{|l|}{$p=0$} & $\mathcal{F}_{2}^{\Lambda }=Q^{\Lambda }$ $%
\mathbf{\alpha }_{_{AdS_{{\small 2}}}}$ & $\mathcal{\tilde{F}}_{6|\Lambda
}=P_{\Lambda }$ $\mathbf{\beta }_{S^{3}}\wedge \mathbf{\beta }_{T^{3}}$ \\
\hline
\multicolumn{1}{|l|}{$p=1$} & $\mathcal{F}_{3}=p_{0}\mathbf{\beta }_{S^{3}}$
& $\mathcal{\tilde{F}}_{5}=q^{0}\left( \mathbf{\alpha }_{_{AdS_{{\small 2}}}}%
{\small \wedge }\mathbf{\beta }_{T^{3}}\right) $ \\ \hline
\end{tabular}%
\end{equation}%
The total effective potential reads, in terms of the dressed central charges
of the black hole/4-brane, the black string/3-brane, as in (\ref{vv}) with
typical solutions at horizon given by $X=\pm \tilde{X},$ $Y=\pm \tilde{Y}$, $%
\sigma _{0}=\frac{-1}{4}\ln \frac{p_{0}}{q_{0}}$. Other solutions of type
eqs(\ref{xxx},\ref{yyy}) may be also written down. Notice also that the
solutions with plus signs describe intersecting attractor involving string,
3-brane, D0- brane and D4- brane; those with minus signs are associated with
the string, 3-brane, ant-D0 and anti D4- brane.

\subsubsection{$AdS_{2}\times S^{2}\times T^{4}$}

The associated field strengths on this geometry read as follows,%
\begin{equation}
\begin{tabular}{l|l|l|}
\cline{2-3}
& p-branes & $\left( 4-p\right) $- branes \\ \hline
\multicolumn{1}{|l|}{$p=0$} & $\mathcal{F}_{2}^{\Lambda }=P^{\Lambda }$ $%
\mathbf{\beta }_{S^{2}}$ & $\mathcal{\tilde{F}}_{6|\Lambda }=Q_{\Lambda }$ $%
\left( \mathbf{\alpha }_{Ads_{2}}\wedge \mathbf{\beta }_{T^{4}}\right) $ \\
\hline
\multicolumn{1}{|l|}{$p=1$} & $\mathcal{F}_{3}=\dsum%
\limits_{k=1}^{4}p_{0}^{k}$ $\left( \mathbf{\beta }_{S^{2}}\mathbf{\wedge
\beta }_{S_{k}^{1}}\right) $ & $\mathcal{\tilde{F}}_{5}=\dsum%
\limits_{l=1}^{4}q_{l}^{0}\varepsilon ^{lijk}$ $\left( \mathbf{\alpha }%
_{Ads_{2}}\mathbf{\wedge \beta }_{S_{i}^{1}}\mathbf{\wedge \beta }%
_{S_{j}^{1}}\mathbf{\wedge \beta }_{S_{k}^{1}}\right) $ \\ \hline
\end{tabular}%
\end{equation}%
Following the same approach we have been using, the effective potential $%
\mathcal{V}_{eff}$ of these black brane configurations is given by,%
\begin{equation}
\begin{tabular}{ll}
$\mathcal{V}_{eff}=$ & $+\dsum\limits_{i,j=1}^{2}\left( X^{i}\delta
_{ij}X^{j}+\tilde{X}_{i}\delta ^{ij}\tilde{X}_{j}\right)
+\dsum\limits_{k=1}^{4}\left( e^{4\sigma }p_{0k}^{2}+e^{-4\sigma
}q_{0k}^{2}\right) $ \\
& $+\dsum\limits_{a,b=1}^{N}\left( Y^{a}\delta _{ab}Y^{b}+\tilde{Y}%
_{a}\delta ^{ab}\tilde{Y}_{b}\right) $%
\end{tabular}%
\end{equation}%
Here also there are various types of solutions describing intersecting
attractors with the moduli space $SO\left( 2\right) \times SO\left( N\right)
$ symmetries broken down to subgroups; a class of them reads as:%
\begin{equation}
\begin{tabular}{llllll}
$\sigma _{0}=\frac{1}{8}\ln \left( \frac{%
q_{01}^{2}+q_{02}^{2}+q_{03}^{2}+q_{04}^{2}}{%
p_{01}^{2}+p_{02}^{2}+p_{03}^{2}+p_{04}^{2}}\right) $ & , & $X=\pm \tilde{X}%
, $ & $Y=\pm \tilde{Y}$ &  & ,%
\end{tabular}%
\end{equation}%
they correspond to the case where $SO\left( 2\right) \times SO\left(
N\right) $ is completely broken. Notice also that for the particular case $%
X=\pm \tilde{X}=0$ and $Y=\pm \tilde{Y}\neq 0$, the moduli space symmetry is
reduced to $SO\left( 2\right) $ and in the case $X=\pm \tilde{X}\neq 0$ and $%
Y=\pm \tilde{Y}=0$, it reduces to $SO\left( N\right) $.

\section{Conclusion}

In this paper, we have studied the attractor mechanism of intersecting black
p-branes in non chiral \emph{8D} supergravity with \emph{16} supercharges.
Actually, this study completes previous results on black attractors in non
chiral \emph{8D} supergravity with \emph{32} supersymmetries \textrm{\cite%
{7A}} and agrees with the results on higher \emph{D}-supergravities obtained
in \textrm{\cite{70}}.\newline
To do so, we have first studied the structure of non chiral \emph{8D }$%
\mathcal{N}=1$ supersymmetric algebra with non trivial central charges $%
Z_{\mu _{1}...\mu _{p}}$. Then we have given the link between these $Z_{\mu
_{1}...\mu _{p}}$s and the fluxes $\int_{S^{2}}\mathcal{F}_{\mu _{1}...\mu
_{p+2}}$ of p-branes; in particular the D- branes of type IIA string on a
compact real surface $\Sigma $ given by eq(\ref{a}). Using group theoretic
method, we have shown that, besides the F-sting and the D0- brane, only the $%
\left( D2/\Sigma \right) $-, $\left( D6/\Sigma \right) $- and $\left(
NS5/\Sigma \right) $-branes wrapping 2-cycles of $\Sigma $ which survive
under compactification; no free D2- nor $\left( D4/\Sigma \right) $-brane
are allowed in non chiral \emph{8D }$\mathcal{N}=1$ supergravity. This
result has been also checked by using a field theoretical method by
determining directly the fields content that follows from \emph{10D} type II
spectrum on $\Sigma $. \newline
We have also studied the attractor mechanism for both standard extremal
black attractors in \emph{8D} supergravity with \emph{16} supercharges as
well as their intersections along the line of \textrm{\cite{70,7A}}. We have
worked out various classes of explicit solutions and shown that they are
completely classified by the $SO\left( N-m\right) $ subgroups of the $%
SO\left( 2\right) \times SO\left( N\right) $ symmetry of the moduli space $%
\frac{SO\left( 2,N\right) }{SO\left( 2\right) \times SO\left( N\right) }%
\times SO\left( 1,1\right) $.


\bigskip

\begin{thebibliography}{99}
\bibitem{1A} Anna Ceresole, Sergio Ferrara,\emph{\ Black Holes and
Attractors in Supergravity}, arXiv:1009.4175,

\bibitem{01A} S. Ferrara, D. Z. Freedman and P. Van Nieuwenhuizen, \emph{%
Progress Toward a Theory of Supergravity}, Phys. Rev. D13 (1976) 3214,%
\newline
S. Deser and B. Zumino, \emph{Consistent Supergravity}, Phys. Lett. 62B
(1976) 335.

\bibitem{01B} E. Cremmer, B. Julia and J. Scherk, \emph{Supergravity theory
in 11 dimensions, }Phys.Lett. B 76 (1978) 409,

\bibitem{03E} S. Ferrara, J.G. Taylor, \emph{Supergravity'81}, \emph{%
Proceedings of the 1st School on Supergravity}, International Centre for
Theoretical Physics, Trieste, Italy,

\bibitem{9A} Sergio Ferrara, Kuniko Hayakawa, Alessio Marrani, \emph{Erice
Lectures on Black Holes and Attractors}, Fortsch.Phys.56:993-1046,2008,
arXiv:0805.2498,

\bibitem{1B} S. Bellucci, S. Ferrara, A. Marrani,\emph{\ Attractors in Black}%
, Fortsch.Phys.56:761-785,2008, arXiv:0805.1310,

\bibitem{1C} Sergio Ferrara, Alessandra Gnecchi, Alessio Marrani, \emph{d=4
Attractors, Effective Horizon Radius and Fake Supergravity},
Phys.Rev.D78:065003,2008, arXiv:0806.3196,

\bibitem{1D} M. J. Duff, S. Ferrara, \emph{Four curious supergravities},
arXiv:1010.3173,

\bibitem{1E} R. Kallosh, \emph{New attractors}, JHEP 0512, 022 (2005),

\bibitem{1F} S. Ferrara and J. M. Maldacena, \emph{Branes, central charges
and U-duality invariant BPS conditions}, Class. Quant. Grav. 15, 749 (1998),
arXiv:9706097,

\bibitem{1G} S. Bellucci, S. Ferrara, A. Shcherbakov, A. Yeranyan, \emph{%
Attractors and first order formalism in five dimensions revisited},
arXiv:1010.3516,

\bibitem{1H} S. Ferrara and R. Kallosh, \emph{Supersymmetry and Attractors},
Phys. Rev. D54 (1996) 1514,arXiv:9602136,

\bibitem{2A} S. Bellucci, S. Ferrara, R. Kallosh, A. Marrani,\emph{\
Extremal Black Hole and Flux Vacua Attractors}, Lect.Notes
Phys.755:115-191,2008,arXiv:0711.4547,

\bibitem{2B} S. Ferrara, G. W. Gibbons and R. Kallosh, \emph{Black Holes and
Critical Points in Moduli Space}, Nucl. Phys. B500, 75 (1997), arXiv:9702103,

\bibitem{2C} Anna Ceresole, Gianguido Dall'Agata, Sergio Ferrara, Armen
Yeranyan,\emph{Universality of the superpotential for d = 4 extremal black
holes\ }, arXiv:0910.2697,

\bibitem{2D} Sergio Ferrara, Renata Kallosh, Andrew Strominger, \emph{N=2
Extremal Black Holes}, Phys.Rev.D52:5412-5416,1995, arXiv:hep-th/9508072,

\bibitem{2E} A. Salam and E. Sezgin, $\emph{D=8}$ $\emph{supergravity}$,
Nucl. Phys. B258, 284 (1985),

\bibitem{2F} Cumrun Vafa, \emph{Lectures on Strings and Dualities},
arXiv:hep-th/9702201,

\bibitem{3A} P. Aschieri, S. Ferrara and B. Zumino, \emph{Duality Rotations
in Nonlinear Electrodynamics and in Extended Supergravity}, Riv. Nuovo Cim.
31, 625 (2009) arXiv:0807.4039,

\bibitem{01C} F. Larsen, The Attractor Mechanism in Five Dimensions,
hep-th/0608191,

\bibitem{01D} R. Kallosh, New Attractors, JHEP 0512 (2005) 022,
hep-th/0510024,

\bibitem{4B} H. Ooguri, A. Strominger, C. Vafa, \emph{Black Hole Attractors
and the Topological String}, Phys.Rev. D70 (2004) 106007, arXiv:0405146,

\bibitem{03A} S. Ferrara, R. Kallosh and A. Strominger, \emph{N=2 extremal
black holes,} Phys. Rev. D52 (1995) 5412,

\bibitem{03B} A. Strominger, \emph{Macroscopic entropy of N=2 extremal black
holes}, Phys. Lett. B383, 39 (1996),

\bibitem{03C} S. Ferrara and R. Kallosh, \emph{Supersymmetry and attractors,
Phys. Rev. D 54}, 1514 (1996),

\bibitem{03D} S. Ferrara and R. Kallosh, \emph{Universality of
supersymmetric attractors}, Phys. Rev. D 54, 1525 (1996)

\bibitem{3B} Anna Ceresole, Sergio Ferrara, and Alessio Marrani, \emph{Small
N=2 Extremal Black Holes in Special Geometry\ }, arXiv:1006.2007,

\bibitem{030B} Lilia Anguelova, Flux Vacua Attractors and Generalized
Compactifications, JHEP 0901:017,2009, arXiv:0806.3820

\bibitem{3C} J.A. Strathdee, \emph{Extended Poincar\'{e} Supersymmetry},
Int. J. Mod. Phys. A2, 273 (1987),

\bibitem{3D} Lilia Anguelova, Finn Larsen, Ross O'Connell, \emph{Heterotic
Flux Attractors}, arXiv:1006.4981,

\bibitem{3E} S. Bellucci, S. Ferrara, A. Shcherbakov, A. Yeranyan, \emph{%
Attractors and first order formalism in five dimensions revisited},
arXiv:1010.3516,

\bibitem{3F} Sergio Ferrara, Alessio Marrani, Emanuele Orazi, \emph{Split
Attractor Flow in N=2 Minimally Coupled Supergravity}, arXiv:1010.2280,

\bibitem{3G} Sergio L. Cacciatori, Dietmar Klemm, \emph{Supersymmetric AdS4
black holes and attractors}, arXiv:0911.4926,

\bibitem{3I} Yun Soo Myung, Yong-Wan Kim, Young-Jai Park, \emph{New
attractor mechanism for spherically symmetric extremal black holes},
Phys.Rev.D76:104045,2007, arXiv:0707.1933,

\bibitem{4A} S. Bellucci, S. Ferrara, M. Gunaydin, A. Marrani, \emph{SAM
Lectures on Extremal Black Holes in d=4 Extended Supergravity},
arXiv:0905.3739,

\bibitem{4C} A. Salam and E. Sezgin, \emph{d=8 Supergravity: Matter Coupling
and Minkowski Compactification }, Physics Letters, 1985,

\bibitem{40C} R. Ahl Laamara, A. Belhaj, L.B. Drissi, E.H. Saidi, \emph{%
Black Holes in Type IIA String on Calabi-Yau Threefolds with Affine ADE
Geometries}, Nucl.Phys.B776:287-326,2007, arXiv:hep-th/0611289,

\bibitem{4D} K. Saraikin and C. Vafa, $\emph{Non}$\emph{-}$\emph{%
supersymmetricBlack}$ $\emph{Holes}$ $\emph{and}$ $\emph{Topological}$ $%
\emph{Strings}$, Class. Quant. Grav. 25, 095007 (2008), hep-th/0703214,

\bibitem{5A} A. Salam and E. Sezgin, \emph{Supergravities in Diverse
Dimensions}, World Scientific, 1989,

\bibitem{5B} L.Andrianopoli, R.D'Auria, S.Ferrara, \emph{Central Extension
of Extended Supergravities in Diverse Dimensions, }hep-th/9608015,

\bibitem{5C} Y. Tanii, \emph{Introduction to Supergravities in Diverse
Dimensions}, arXiv:hep-th/9802138,

\bibitem{5D} A. Belhaj, L. B. Drissi, E. H. Saidi, A. Segui, \emph{N=2
Supersymmetric Black Attractors in Six and Seven Dimensions},
Nucl.Phys.B796:521-580,2008, arXiv:0709.0398,

\bibitem{5E} Sergio Ferrara, Murat Gunaydin, \emph{Orbits and Attractors for
N=2 Maxwell-Einstein Supergravity Theories in Five Dimensions},
Nucl.Phys.B759:1-19,2006, arXiv:hep-th/0606108,

\bibitem{6A} Lilia Anguelova, \emph{Flux Vacua Attractors and Generalized
Compactifications}, JHEP 0901:017,2009, arXiv:0806.3820,

\bibitem{6B} T.G. Pugh, E. Sezgin, K.S. Stelle, \emph{D=7 / D=6 Heterotic
Supergravity with Gauged R-Symmetry},arXiv:1008.0726,

\bibitem{6C} E. Witten, $\emph{String}$ $\emph{Theory}$ $\emph{Dynamics}$ $%
\emph{in}$ $\emph{Various}$ $\emph{Dimensions}$, Nucl. Phys. B 443(1995)184,

\bibitem{6D} El Hassan Saidi, \emph{BPS and non BPS 7D Black Attractors in
M-Theory on K3}, arXiv:0802.0583,

\bibitem{06D} El Hassan Saidi, \emph{On Black Hole Effective Potential in
6D/7D N=2 Supergravity}, Nucl.Phys.B803:235-276,2008, arXiv:0803.0827,

\bibitem{7B} D.Z. Freedman, P. van Nieuwenhuizen, S. Ferrara, \emph{Progress
Toward a Theory of Supergravity}, Phys.Rev.D13:3214-3218,1976,

\bibitem{7C} J. M. Maldacena, A. Strominger and E. Witten, \emph{Black hole
entropy in M-theory}, JHEP 9712, 002 (1997),

\bibitem{7D} A. Strominger and C. Vafa, \emph{Microscopic Origin of the
Bekenstein-Hawking Entropy}, Phys. Lett. B 379, 99 (1996),

\bibitem{8A} A. Sen,\emph{\ Black hole entropy function and the attractor
mechanism in higher derivative gravity} , JHEP 0509, 038 (2005),
arXiv:0506177,

\bibitem{8B} A. Dabholkar, \emph{Black hole entropy and attractors}, Class.
Quant. Grav. 23 (2006) 957-980,

\bibitem{8C} S. Ferrara and R. Kallosh,\emph{\ Universality of
Supersymmetric Attractors}, Phys. Rev.D54, 1525 (1996), arXiv:9603090,

\bibitem{8D} M. Gunaydin, G. Sierra and P. K. Townsend,\emph{\ Exceptional
Supergravity Theories and the Magic Square}, Phys. Lett. B133, 72 (1983),

\bibitem{ss} E.H Saidi, A. Segui, \emph{Entropy of Pairs of Dual Attractors
in 6D/7D}, J. High Energy Phys. JHEP07(2008)128, arXiv:0803.2945,

\bibitem{70} S. Ferrara, A. Marrani, J. F. Morales, H. Samtleben, \emph{%
Intersecting Attractor},\emph{\ }Phys.Rev.D79:065031,2009, arXiv:0812.0050,

\bibitem{7A} L. B. Drissi, F. Z. Hassani, H. Jehjouh, E. H. Saidi, \emph{%
Extremal Black Attractors in 8D Maximal Supergravity},
PhysRevD.81.105030,2010, arXiv:1008.2689,

\bibitem{100} El Hassan Saidi, \emph{On Black Attractors in}\textbf{\ \emph{%
8D }}\emph{and Heterotic/Type IIA Duality}, CPM-10-01,

\bibitem{10A} L. Andrianopoli, R. D'Auria, S. Ferrara, P. Fr\'{e}, M.
Trigiante, R--R Scalars, \emph{U--Duality and Solvable Lie Algebras,Nucl.Phys%
}. B496 (1997) 617-629, arXiv:hep-th/9611014,

\bibitem{11A} J.D. Edelstein, A. Paredes, A.V. Ramallo,\emph{\ Wrapped
branes with fluxes in 8d gauged supergravity},JHEP 0212 (2002)
075,arXiv:hep-th/0207127.
\end{thebibliography}
\end{document}